\documentclass[a4paper, 12pt]{article}
\usepackage{graphicx} 
\usepackage{amsmath}
\usepackage{amssymb}
\usepackage{units}
\usepackage{color}
\usepackage{todonotes}
\usepackage[T1]{fontenc}
\DeclareUnicodeCharacter{2212}{-}

\usepackage{tikz-feynman}
\tikzfeynmanset{compat=1.1.0}
\usepackage{feynmp}
\usepackage{jheppub} 

\title{Quasi-Dirac Heavy Neutral Leptons in the Left-Right Symmetric Model}
\author{Oleksii~Mikulenko}
\emailAdd{mikulenko@lorentz.leidenuniv.nl}
\affiliation{Instituut-Lorentz, Leiden University, Niels Bohrweg 2, 2333 CA Leiden, The Netherlands}

\begin{document}

\abstract{
    We discuss the phenomenology of a pair of degenerate GeV-scale Heavy Neutral Leptons within the Left-Right Symmetric Model (LRSM) framework,  with the third fermion serving as a dark matter candidate. We highlight the potential of the recently approved SHiP experiment to test the existence of the light DM species, and the signatures of lepton number violation as a possible experimental probe of the model in various experiments. Our findings include concrete predictions, in some part of the model's parameter space, for the effective right-handed couplings $(V^{R}_{e})^2:(V^{R}_{\mu})^2:(V^{R}_{\tau})^2 = 0.16:0.47:0.38$ (normal neutrino hierarchy), $0.489 : 0.22 : 0.30$ (inverted hierarchy) of the degenerate pair. 

    }
    
\maketitle

\section{Introduction.}
The Standard Model (SM) has achieved remarkable success in describing physics at energies up to the electroweak scale, both in collider experiments and in the universe. A few observed phenomena, however, still lack an explanation in the SM: the existence of nonluminous, nonbaryonic dark matter that permeates the Universe; masses of neutrinos, manifesting in the oscillations between different flavors; and the striking asymmetry between the matter and antimatter in our Universe. A minimalistic and natural scenario, capable of handling all the aforementioned beyond the SM (BSM) problems, consists of introducing right-handed counterparts of the SM neutrinos, i.e. sterile neutrinos~\cite{Minkowski:1977sc,Mohapatra:1979ia,Mohapatra:1980yp,Schechter:1980gr}. As the name indicates, these new particles are not charged under any of the SM gauge groups, may have Majorana mass, and are an example of more general Heavy Neutral Leptons (HNL), see~\cite{Mohapatra:2005wg,Abdullahi:2022jlv} for a review. Despite being sterile, the new particles possess a feeble weak-like interaction as a consequence of mixing with active neutrinos. The interaction is suppressed by the small mixing angle $U\ll 1$ and opens up the opportunity to search for these elusive particles at future accelerator experiments, such as SHiP~\cite{SHiP:2015vad,SHiP:2018xqw}, DUNE~\cite{DUNE:2015lol,Ballett:2019bgd,Krasnov:2019kdc}, LHC-based experiments~\cite{Kling:2018wct,Boyarsky:2021moj,Curtin:2018mvb,Dercks:2018wum,Hirsch:2020klk,Aielli:2019ivi,Feng:2022inv,Ovchynnikov:2022its}, and future colliders~\cite{Antusch:2016ejd,Blondel:2022qqo, Chrzaszcz:2020emg,Mekala:2022cmm, Boyarsky:2022epg, Li:2023tbx, Kwok:2023dck}.  In addition to this, HNLs can be constrained with astrophysics and cosmology~\cite{Dolgov:2000jw, Sabti:2020yrt, Boyarsky:2020dzc,Vincent:2014rja,Poulin:2016anj,Syvolap:2019dat,Rembiasz:2018lok,Suliga:2019bsq,Coloma:2017ppo,Gustafson:2022rsz, Fischer:2022zwu} and indirect searches such as electroweak precision measurements~\cite{Antusch:2014woa, Fernandez-Martinez:2016lgt, Chrzaszcz:2019inj} and charge lepton number violation~\cite{Bernstein:2013hba,Calibbi:2017uvl,Urquia-Calderon:2022ufc}. 

The future experiments envisioned can observe HNLs with couplings orders of magnitude beyond the current limits. Consequently, there is a potential for abundant signal yield, which would facilitate scrutinizing the properties of the newly seen particles~\cite{Mikulenko:2023iqq, Mikulenko:2023olf}. A natural question then arises: what is the landscape of models that could, in principle, lead to such an observed signal, and how can we differentiate between these models? This question is the main motivation of the presented work.

The neutrino Minimal Standard Model ($\nu$MSM)~\cite{Asaka:2005an,Asaka:2005pn} introduces one keV-scale sterile neutrino $N_1$ and two heavier HNLs $N_2$, $N_3$ with masses at or above the GeV scale. For the naming convention, we will denote the heavy species that can be probed by direct searches as HNLs ($N_2$, $N_3$ in this case). The sterile neutrino $N_1$ serves as a warm dark matter candidate~\cite{Drewes:2016upu, Boyarsky:2018tvu} and has a somewhat narrow range of possible masses around the keV scale, bounded from below by the Tremaine-Gunn bound~\cite{Tremaine:1979we, Boyarsky:2008ju} and from above by the X-ray emission constraints from DM-dominated objects~\cite{Abazajian:2001vt}. The two HNLs lead to two massive neutrino states via the so-called type-I seesaw mechanism, leaving the lightest active neutrino effectively massless.\footnote{Neglecting the contribution of the dark matter $N_1$, which is tiny due to the constraints on the particle's lifetime.} The naive scale of the HNL mixing angles is then defined by the seesaw limit $U^2_\text{seesaw} = \sqrt{\Delta m^2_\text{atm.}}/m_N$, with $\Delta m^2_\text{atm.} =(\unit[50]{meV})^2$ being the mass scale of the active neutrinos. This scale is too small to be reached by the mentioned future experiments. An attractive theoretical scenario is to consider the HNL pair as a \textit{quasi-Dirac} fermion, which alleviates some problems. First, an approximate Dirac nature translates into a cancellation between the two seesaw contributions: the mixing angles of HNLs may be large, within the reach of the near-future experiments, without spoiling active neutrino masses. Second, there is limited freedom in the relations between the coupling constants to different lepton flavors, making the model less generic and more easily verifiable. Finally, oscillations between the two GeV degenerate HNLs may enhance the efficiency of leptogenesis~\cite{Akhmedov:1998qx,Canetti:2010aw,Canetti:2012vf,Klaric:2020phc,Klaric:2021cpi}, a possible mechanism for the generation of the baryon asymmetry in the Universe~\cite{Fukugita:1986hr,Davidson:2008bu} which is otherwise viable only for HNL masses (much) above the electroweak scale. Ultimately, the model deals with three BSM phenomena and is a very attractive target for searches.

Beyond the minimal model, HNLs can possess additional interactions, which may affect searches for such particles~\cite{Magill:2018jla,Ovchynnikov:2023wgg,Ovchynnikov:2022rqj,deGiorgi:2022oks,Abdullahi:2023gdj,Fernandez-Martinez:2023phj}. One model that naturally leads to such nonminimal interactions is the Left-Right Symmetric Model (LRSM)~\cite{Pati:1974yy,Mohapatra:1979ia,Mohapatra:1980yp}, attempting to restore the symmetry between the left and right particles. The model extends the SM with a new $SU(2)_R$ gauge symmetry and equips HNLs, the right-handed neutrino siblings in this case, with interactions through the right-handed charged current. The new symmetry is spontaneously broken at a scale higher than the electroweak scale, and the processes with the right-handed analog $W_R$ of the $W$-boson are suppressed by the large mass of the mediator $m_{W_R}\gg m_W$. The constraints on the scale of new physics come from direct searches~\cite{CMS:2018hff,ATLAS:2019lsy,CMS:2021dzb,ATLAS:2023cjo, Cottin:2019drg}, meson precision measurements~\cite{Bertolini:2014sua}, and neutrinoless double beta decay~\cite{Tello:2010am, Gluza:2016qqv}. Current limits on the mass of the new boson lie generally around or below $\unit[5]{TeV}$. 

In this work, we embed the $\nu$MSM-motivated case of one sterile neutrino and two HNLs in the quasi-Dirac limit into the minimal Left-Right Symmetric Model. We study the flavor structure of the HNL couplings and analyze the additional signatures arising from the new interactions, based solely on the seesaw relation between the couplings and the active neutrino masses. Specifically, we found the analytic form of the leptonic analog $V^R$ of the CKM matrix, in the limit of type-I seesaw, depending on two unknown parameters. In this case and for some experimental setups, the HNL pair can be described as a single particle with effective coupling \textit{uniquely fixed} by the experimentally measured properties of active neutrinos. In addition, we compute the corrections to the matrices of left and right-handed couplings arising from the full type-II seesaw relation.

It must be stressed that the considered embedding has a few problems. From the phenomenological perspective, we treat the $\nu$MSM as a limit of the LRSM, in which the energy scale of new gauge bosons is too high to affect experimental searches. The DM candidate $N_1$ can still remain cosmologically stable independent of the new interaction if it is lighter than the electron, i.e. the decay through right-handed current is kinematically forbidden. However, the inclusion of gauge interactions with the new scale in the $\unit[10]{TeV}-\unit[100]{TeV}$ range (within the reach of collider experiments) introduces an additional means of thermalization of particles in the early Universe. This imposes severe obstacles to the model's capability to account for both baryon asymmetry~\cite{ Frere:2008ct,Bezrukov:2012as,BhupalDev:2014hro,Dhuria:2015cfa} and dark matter~\cite{Bezrukov:2009th}. Thus, some additional mechanisms are needed for the model to be viable. The thermal overproduction of dark matter may be avoided with a period of early matter domination by some heavy particles, diluting the abundance of hot relics~\cite{Nemevsek:2012cd,Borah:2017hgt}, although the realization of such a scenario within the experimentally interesting region is challenging~\cite{Nemevsek:2022anh, Nemevsek:2023yjl}.

The structure of the paper is as follows. Section 2 describes the phenomenology of HNLs with combined left and right-current interactions. In Section 3, we constrain the flavor structure of the left and right coupling constants from the initial LRSM lagrangian. Section 4 discusses various interesting experimental signatures of the considered model, specifically the lepton number violation signatures and dark matter search at the SHiP experiment. We conclude in Section 5.

\section{Phenomenology of HNLs below electroweak scale}
\label{sec:pheno}

In this section, we summarize the phenomenology of a unified left-plus-right interaction of HNLs that will be relevant to us. Without going into details, we assume that the gauge symmetry contains two subgroups $SU_L(2)$ and $SU_R(2)$ that are coupled correspondingly to left and right fermions and get spontaneously broken at different scales. The right-handed scale is higher, making the associated gauge bosons much heavier than the SM $W$, $Z$-bosons. Right-handed HNLs are connected to the light SM fermions via the charged boson $W_R$ and through the Yukawa coupling to neutrinos (mixing). For a right-handed Weyl spinor $N_I$, the Lagrangian contains the following interactions through left (LH) and right-handed (RH) currents
\begin{align*}
    \mathcal L_L \supset \;&\frac{g}{\sqrt{2}} \bar l \,N_I^c W + \frac{g}{2 \cos \theta_W} \bar \nu\, N_I Z    \\
    \mathcal L_R \supset \;& \frac{g}{\sqrt{2}} \bar l\,  N_I W_R 
\end{align*}
Here, $W$, $Z$ are the SM gauge bosons, $W_R$ is the charged boson of the $SU_R(2)$, and $\alpha = e,\mu,\tau$ are leptonic flavors. 
In principle, left and right bosons can mix with each other: thus, $W$ may mediate the RH interactions as well. We do not consider such interactions, see Sec.~\ref{sec:typeInonzerob} for further details. The mixing between $Z$ and its heavier counterpart $Z_R$ is relevant in the $Z$-pole experiments and is considered separately in Sec.~\ref{sec:FCCee}. The interactions mediated by the scalar sector of the model are not considered here: we expect them to be subdominant because of an additional suppression from the Yukawa couplings of particles in the considered GeV-scale.  A review of all interactions in LRSM can be found in~\cite{Roitgrund:2014zka}.

For HNLs below the electroweak scale, the interactions unify into a generalized Fermi interaction with both left and right currents:
\begin{align}
    \mathcal L \supset\quad& \theta^L_{\alpha I} \frac{G_F}{\sqrt{2}}  \bar l_\alpha N^c_I \times [\bar\nu_\beta l_{\beta} + V^\text{CKM}_{ij} \bar u_{i,L}  d_{j,L}]
    + \theta^L_{\alpha I} \frac{G_F}{\sqrt{2}} \bar \nu_\alpha N^c_I J_Z && \bigg| \,\,\text{LH}
    \nonumber
    \\
    + &\theta^R_{\alpha I} \frac{G_F}{\sqrt{2}} \bar l_{\alpha} N_I \times \bigg[\tilde V^R_{J \beta} \bar N_{J} l_{\beta}  + V^{R,\text{CKM}}_{ij} u_{i,R} d_{i,R}\bigg] & & \bigg| \,\,\text{RH}
    \nonumber
    \\    
    \label{eq:effFermi}
    + &\text{h.c.}
\end{align}
containing some phenomenological couplings, with $G_F$ being the Fermi constant and $J_Z$ being the SM neutral current. Both types of interactions are suppressed by small parameters\footnote{We use the upper and lower-index notation $\theta^L = \theta_L$, $V^R = V_R$, etc., interchangeably} $\theta_{L,R}\ll 1$. To separate the suppression scale and the flavor structure, it is useful to reparametrize them in the form
\begin{equation*}
\label{eq:theta_definition}
    \theta^L_{\alpha I} \equiv U^L_{I} V^L_{\alpha I}, \qquad  \theta^R_{\alpha I} \equiv U^R_I V^R_{\alpha I}, \qquad \sum_\alpha |V^{R/L}_{\alpha I}|^2 = 1.
\end{equation*}

The Lagrangian~\eqref{eq:effFermi} may be viewed as an effective theory with a set of arbitrary couplings. For some experimental probes, the constraints on the $\theta_L$ can be recasted into the equivalent bounds for $\theta_R$. For example, this can be applied to the $0\nu\beta\beta$ bounds~\cite{Mitra:2011qr}; one has to be careful, however, with the possible interference effects~\cite{Gluza:2016qqv}.  The Left-Right symmetric model imposes the following constraints. First, there are 3 HNL species, and the right-handed suppression scale $U^R_I = m^2_{W}/m^2_{W_R}$\footnote{Or, generally, $g_R m_W^2/g_L m^2_{W_R}$, if one does not impose the equality of the left and right gauge couplings $g_L \neq g_R$.} is equal for all species. Second, the unspecified two-HNL interaction couplings $\tilde V^R_{J\beta}$ must be equal to $(V^{R}_{\beta J})^*$. Finally, the right-handed CKM matrix $V^{R,\text{CKM}}$ is approximately equal to the standard CKM matrix, up to the corrections that are bounded from above by $2m_b/m_t \approx0.05$~\cite{Senjanovic:2015yea}. In this work, we assume that these two CKM matrices are equal.

\begin{figure}
    \centering

         \begin{minipage}[h!]{0.4\textwidth}
        \centering

\begin{tikzpicture}
  \begin{feynman}
    \vertex (i) {$N_I$};
    \vertex[right=2cm of i] (v1) ;
    \vertex[above right=0.6cm of v1] (a) ;
    \vertex[left=0.2cm of a] {$\theta^L_{\alpha I}$};
    \vertex[above right=2cm of v1] (o1) {$l_\alpha$};
    \vertex[below right=1.5cm of v1] (v2);
    \vertex[above right=1.5cm of v2] (o2) {$\nu_\beta/u_i$};
    \vertex[below right=1.5cm of v2] (o3) {$\bar l_\beta/\bar d_j$};

    \diagram* {
      {[edges=fermion]
        (i) -- (v1) -- (o1),
        (o3) -- (v2) -- (o2),
      },
      (v2) -- [boson, edge label=\(W\)] (v1),

    };
  \end{feynman}
\end{tikzpicture}
\begin{tikzpicture}
  \begin{feynman}
    \vertex (i) {$N_I$};
    \vertex[right=2cm of i] (v1) ;
    \vertex[above right=0.6cm of v1] (a) ;
    \vertex[left=0.2cm of a] {$\theta^L_{\alpha I}$};
    \vertex[above right=2cm of v1] (o1) {$\nu_\alpha$};
    \vertex[below right=1.5cm of v1] (v2);
    \vertex[above right=1.5cm of v2] (o2) {$f$};
    \vertex[below right=1.5cm of v2] (o3) {$\bar f$};

    \diagram* {
      {[edges=fermion]
        (i) -- (v1) -- (o1),
        (o3) -- (v2) -- (o2),
      },
      (v2) -- [boson, edge label=\(Z\)] (v1),

    };
  \end{feynman}
\end{tikzpicture}    
    
\end{minipage}~
\begin{minipage}[h!]{0.4\textwidth}
        \centering
\begin{tikzpicture}
  \begin{feynman}
    \vertex (i) {$N_I$};
    \vertex[right=2cm of i] (v1) ;
    \vertex[above right=0.6cm of v1] (a); 
    \vertex[left=0.2cm of a] {$V^R_{\alpha I}$};
    \vertex[above right=2cm of v1] (o1) {$l_\alpha$};
    \vertex[below right=1.5cm of v1] (v2);
    \vertex[above right=1.5cm of v2] (o2) {$u_i$};
    \vertex[below right=1.5cm of v2] (o3) {$\bar d_j$};

    \diagram* {
      {[edges=fermion]
        (i) -- (v1) -- (o1),
        (o3) -- (v2) -- (o2),
      },
      (v2) -- [boson, edge label=\(W_R\)] (v1),

    };
  \end{feynman}
\end{tikzpicture}

\begin{tikzpicture}
  \begin{feynman}
    \vertex (i) {$N_I$};
    \vertex[right=2cm of i] (v1) ;
    \vertex[above right=0.6cm of v1] (a); 
    \vertex[left=0.2cm of a] {$V^R_{\alpha I}$};
    \vertex[above right=2cm of v1] (o1) {$l_\alpha$};
    \vertex[below right=1.5cm of v1] (v2);
    \vertex[left=0.9cm of v2] (a2);
    \vertex[below right=0.1cm of a2] {$V^{*R}_{\beta J}$};
    \vertex[above right=1.5cm of v2] (o2) {$N_J$};
    \vertex[below right=1.5cm of v2] (o3) {$\bar l_\beta$};

    \diagram* {
      {[edges=fermion]
        (i) -- (v1) -- (o1),
        (o3) -- (v2) -- (o2),
      },
      (v2) -- [boson, edge label=\(W_R\)] (v1),

    };
  \end{feynman}
\end{tikzpicture}

\end{minipage}

    \caption{Diagrams of the decay of an HNL. \textit{Left:} decays through mixing with active neutrinos. The suppression comes from the small mixing angle $\theta^L_{\alpha I}$. \textit{Right:} decays involving the right-handed interaction. These diagrams are suppressed by $\theta^R_{\alpha I} = V^R_{\alpha I} (m_W/m_{W_R})^2$ at energies below the $W_R$ mass. The bottom-right diagrams have two HNLs while being suppressed by $\theta^R$ only once.}
    \label{fig:HNL decays}
\end{figure}
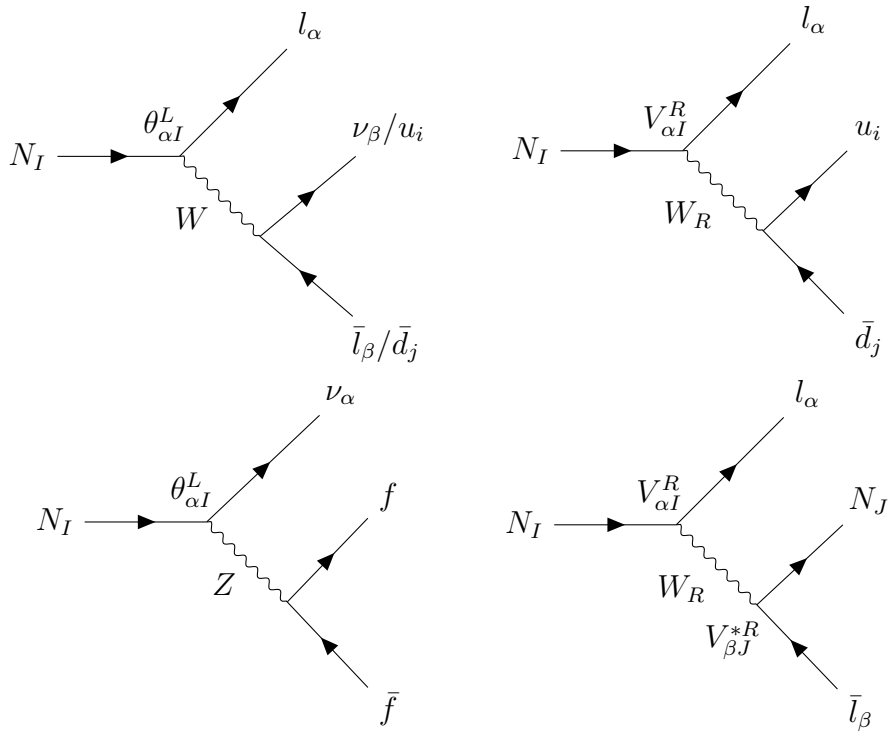

The phenomenology of HNLs with only mixing interactions, $U^R = 0$, has been reviewed in~\cite{Bondarenko:2018ptm}. The main qualitative difference that arises from the inclusion of right-handed interactions is \textit{the appearance of the interaction between different HNL species}. In the minimal case, these would be suppressed by the small left-handed couplings twice, once per each HNL involved. This opens up an opportunity to probe HNLs which would otherwise be inaccessible. 

\begin{figure}[h!]
    \centering

         \begin{minipage}[h!]{0.4\textwidth}
        \centering

\begin{tikzpicture}
  \begin{feynman}
    \vertex (i) {$N_I$};
    \vertex[right=1.5cm of i] (v1) ;
    \vertex[above right=0.6cm of v1] (a); 
    \vertex[left=0.2cm of a] {$V^R_{\alpha I}$};
    \vertex[above right=1.5cm of v1] (o1) {$l_\alpha$};
    \vertex[right=1.5cm of v1] (v2);
    \vertex[above=0.cm of v2] (a2) {$V^{*R}_{\beta J}$};
    \vertex[right=1.5cm of v2] (v3);
    \vertex[above=0.5cm of v3] (o2) {$N_J$};
    \vertex[below=0.5cm of v3] (o3) {$\bar l_\beta$};

    \diagram* {
      {[edges=fermion]
        (i) -- (v1) -- (o1),
        (o3) -- (v2) -- (o2),
      },
      (v2) -- [boson, edge label=\(W_R\)] (v1),

    };
  \end{feynman}
\end{tikzpicture}
    
\end{minipage}~
\begin{minipage}[h!]{0.4\textwidth}
        \centering
\begin{tikzpicture}
  \begin{feynman}
    \vertex (i) {$N_I$};
    \vertex[right=1.5cm of i] (v1) ;
    \vertex[above right=0.6cm of v1] (a); 
    \vertex[left=0.2cm of a] {$V^{*R}_{\beta I}$};
    \vertex[above right=1.5cm of v1] (o1) {$\bar l_\beta$};
    \vertex[right=1.5cm of v1] (v2);
    \vertex[above=0.1cm of v2] (a2) {$V^{R}_{\alpha J}$};
    \vertex[right=1.5cm of v2] (v3);
    \vertex[above=0.5cm of v3] (o2) {$N_J$};
    \vertex[below=0.5cm of v3] (o3) {$l_\alpha$};

    \diagram* {
      {[edges=fermion]
        (o1) -- (v1) -- (i),
        (o2) -- (v2) -- (o3),
      },
      (v2) -- [boson, edge label=\(W_R\)] (v1),

    };
  \end{feynman}
\end{tikzpicture}

\end{minipage}

    \caption{Two interfering diagrams, contributing to the $N_I \to N_J l_\alpha \bar l_\beta$ process. When $N_J$ is produced being ultrarelativistic, the diagrams decouple and correspond to the emission of a specific helicity state.}
    \label{fig:NtoNll}
\end{figure}
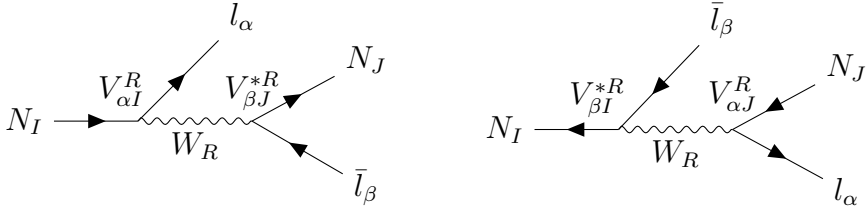

Below, we list the relevant production and decay channels for the model.
\begin{enumerate}
    \item[\textbf{Production.}]
    
For beam-dump experiments, the main HNL production channel is decays of copiously produced mesons. Since strong interactions respect parity, the meson form-factors remain the same regardless of whether the decay happens through the left or the right quark current. The branching ratio for decay of a meson $X$ has the following scalling
    \begin{equation*}
        \text{Br}(X\to Y l_\alpha N_I) = (|\theta^L_{\alpha I}|^2 + |\theta^R_{\alpha I}|^2) \text{Br}_{\text{norm.}}(X\to Yl_\alpha N_\alpha)
    \end{equation*}
    where $(\text{norm.})$ with $N_\alpha$ stands for a quantity, normalized to the $\theta^L_{\beta} = \delta_{\alpha \beta}$ with all the other couplings set to zero. These can be found in~\cite{Bondarenko:2018ptm}.  
    
    At LHC and future high-energy colliders~\cite{Blondel:2022qqo,Curtin:2018mvb,Aielli:2019ivi,Hirsch:2020klk}, decays of $W$, $Z$ (and $H$) become relevant. These happen only through the left current
    \begin{equation*}
        \text{Br}(W,Z,H\to \nu_\alpha N_I) = |\theta^L_{\alpha I}|^2 \text{Br}_\text{norm.}(W,Z,H\to \nu_\alpha N_\alpha)
    \end{equation*}

    For even higher-energy searches, with HNL mass above the electroweak scale, a more case-to-case basis is needed.  

\item[\textbf{Decays.}]
\begin{enumerate}
    \item Hadronic decays through charge current are mediated by both interactions and have equal scaling:
    \begin{equation*}
        \Gamma(N_I \to l_\alpha h^+) = (|\theta^L_{\alpha I}|^2 + |\theta^R_{\alpha I}|^2) \Gamma_{\text{norm.}}(N_\alpha \to l_\alpha h^+),
    \end{equation*}
under our assumption of the equality of left and right-handed CKM matrices. These decays result in a fully detectable state of a charged lepton and hadrons.
    \item Decays into neutral hadrons and neutrino is only mediated by left interaction
    \begin{equation*}
        \Gamma(N_I \to \nu_\alpha h^0) = |\theta^L_{\alpha I}|^2  \Gamma_{\text{norm.}}(N_\alpha \to \nu_\alpha h^0) 
    \end{equation*}
    \item Purely leptonic decays $\Gamma(N_I\to \nu l \bar l \text{ or } 3\nu) \propto |\theta^L|^2$ are mediated by the left mixing only as well
    
    \item Finally, the additional channel of decay into a lighter HNLs $N_I \to N_J l_\alpha \bar l_\beta$ is suppressed only once despite involving two HNL species. There are two diagrams for this process (Fig.~\ref{fig:NtoNll}) that, in general, interfere with each. However, if the lighter species is substantially lighter than the decaying HNL, the two diagrams decouple and start to correspond to the emission of $N_J$ with specific helicity: positive (negative) when $N_J$ acts as a particle (antiparticle). In this limit, the total decay width reads
    \begin{equation*}
        \Gamma(N_I \to N_J l_\alpha \bar l_\beta) = U_R^2 \left(|V^R_{\alpha I}|^2 |V^R_{\beta J}|^2 + |V^R_{\beta I}|^2 |V^R_{\alpha J}|^2\right) \Gamma_\text{norm.}(N_\alpha \overset{W}{\to} l_\alpha \nu_\beta \bar l_\beta)
    \end{equation*}
    The decay is mediated only by $W$. The full expression for the decay width with nonvanishing $m_{N_J}$ is given in App.~\ref{app:decayIntoDM}.
\end{enumerate}
    

    
\end{enumerate}

\section{Flavor structure from the seesaw relation.}
In this section, we derive the numerical values of the $V^L$, $V^R$ matrices from the seesaw relation, which determines the properties of active neutrinos.

\subsection{LRSM Lagrangian}
The leptonic Yukawa sector of the LRSM model is~\cite{Senjanovic:2016vxw}:
\begin{equation*}
    \mathcal L \supset \bar L_\alpha ([Y_e]_{\alpha\beta} \Phi - [Y_\nu]_{\alpha \beta} \sigma_2 \Phi^*\sigma_2) R_\beta + \bar L_\alpha^c [Y_1]_{\alpha \beta} i\sigma_2\Delta_L L_\beta + \bar R_\alpha^c [Y_2]_{\alpha \beta} i\sigma_2\Delta_R R_\beta + \text{h.c.}
\end{equation*}
where $L_\alpha = (\nu, e_L)_\alpha$, $R_{\alpha} = (N, e_R)_\alpha$ are left and right lepton doublets, and $\Phi$, $\Delta_L$, $\Delta_R$ are scalar fields that acquire the following vacuum expectation values:
\begin{equation*}
    \Phi \to v\,\text{diag}(\cos b, -\sin b \,e^{-i a})\qquad \Delta_{L,R} \to \begin{pmatrix}
        0 & 0 \\ v_{L,R} & 0
    \end{pmatrix} 
\end{equation*}

The generalized parity symmetry $\mathcal P$ that exchanges the left and right fields $L \leftrightarrow R$, $\Delta_L \leftrightarrow \Delta_R$ implies the following relations:
\begin{equation*}
    Y^\dagger_e = Y_e, \quad Y^\dagger_\nu = Y_\nu, \quad Y_1 = Y_2 
\end{equation*}

Choosing the initial flavor basis that diagonalizes the charge lepton mass matrix $m_l$ after the symmetry breaking, the Lagrangian takes the form
\begin{equation*}
    \mathcal L \supset (m^\text{diag}_l)_{\alpha \beta} \bar l_\alpha l_\beta - v[\tilde Y_\nu]_{\alpha \beta} \bar \nu_\alpha N_\beta + \frac{v_L}{v_R} M_{\alpha \beta} \bar \nu^c_\alpha \nu_\beta + M_{\alpha \beta} \bar N^c_\alpha N_\beta
\end{equation*}
with the Dirac mass matrix:
\begin{equation*}
    v \tilde Y_\nu \approx v Y_\nu \frac{\cos^2 b- e^{2ia}\sin^2 b}{\cos b} - m^\text{diag}_l e^{ia}\tan b    
\end{equation*}

The neutrino mass matrix becomes a sum of type-I and type-II seesaw contributions:
\begin{equation*}
    m_\nu = - v^2 \tilde Y_\nu M^{-1} \tilde Y_\nu^T + \frac{v_L}{v_R} M 
\end{equation*}

The diagonalization of the HNL mass matrix $M_{\alpha\beta}$ is achieved via a $V_R$ unitary rotation:
\begin{equation*}
    N_\alpha = [V_R]_{\alpha i} \tilde N_I, \qquad M \equiv V^*_{R} m^\text{diag}_N V_R^\dagger, \qquad \tilde Y_\nu \equiv V_R Y V_R^\dagger 
\end{equation*}
and the resulting seesaw relation reads:
\begin{equation}
\label{eq:full_seesaw}
    U_{\text{PMNS}}^* m_\nu^\text{diag} U_{\text{PMNS}}^\dagger = - v^2 V_R Y  [{m^\text{diag}_N}]^{-1} Y^T V_R^T +  \frac{v_L}{v_R} V^*_R m^\text{diag}_N V^\dagger_R
\end{equation}
where the neutrino mass matrix is written explicitly diagonalized with the PMNS matrix.

The identification of the effective couplings that enter Eq.~\eqref{eq:effFermi} is straightforward and given by
\begin{equation*}
    \theta^R_{\alpha I} = \frac{m_{W_L}^2}{m_{W_R}^2} [V^R]_{\alpha I}, \qquad \theta^L_{\alpha I} = \frac{i v}{m_{N_I}} [V^R Y]_{\alpha I}
\end{equation*}

\subsection{Approximate symmetry limit in type-I seesaw, $b=0$}

\label{sec:TypeIcase}
We start with the simplest scenario of the type-I seesaw, with no CP-violation $b=0$ and no type-II contribution $v_L/v_R=0$. To remind the reader, we are seeking a solution that involves only two degenerate HNLs whose mixing angles are allowed to exceed the standard seesaw bound significantly. For this, one can typically employ the Casas-Ibarra parametrization~\cite{Casas:2001sr}. However, this approach faces the problem of ensuring that the Yukawa matrix $Y_\nu$ is hermitian (with $b=0$), a requirement that is not automatically imposed by the parametrization. An analysis discussing the reconstruction of new physics properties from the seesaw relation can be found in~\cite{Nemevsek:2012iq, Senjanovic:2016vxw,Senjanovic:2018xtu, Senjanovic:2019moe}. In this work, we employ a different approach to emphasize the quasi-Dirac nature of the HNL pair.

We start by realizing an exact lepton symmetry that makes the mixing angles completely independent from the neutrino masses.
The exact lepton symmetry that yields vanishing $m_\nu$ and hermitian $Y_\nu$ is achieved by choosing:
\begin{equation}
\label{eq:full_symmetry}
    m_N^\text{diag} = m_N\begin{pmatrix} 0 & 0 & 0 \\ 0 & 1 & 0 \\ 0 & 0 & 1 \end{pmatrix},\qquad Y = y \begin{pmatrix} 0 & 0 & 0 \\ 0 & 1 & -i \\ 0 &  i & 1   \end{pmatrix}
\end{equation}
The other choice with the matrix $Y\to Y^T$ simply corresponds to the redefinition $N_2\leftrightarrow N_3$.

The left-mixing matrix takes the form
\begin{equation*}
    \theta_{\alpha I} = \frac{v}{m_N} (V_R Y)_{\alpha I} = \frac{\sqrt{2} y v}{m_N} V_L 
\end{equation*}
\begin{equation*}
    V_L = \frac{1}{\sqrt{2}}     \begin{pmatrix} 
        0 & \quad V^R_{12} + i V^R_{13} & \quad-i (V^R_{12} + iV^R_{13})\\
        0 & \quad V^R_{22} + i V^R_{23} & \quad-i (V^R_{22} + iV^R_{23})\\
        0 & \quad V^R_{32} + i V^R_{33} & \quad-i (V^R_{32} + iV^R_{33})
    \end{pmatrix}
\end{equation*}
Its structure is completely determined by the right-handed matrix $V^R$.

Now, let us try to violate the exact symmetry in the type-I seesaw terms by introducing small deviations that still respect the hermitian form of $Y$:
\begin{equation*}
    m_N^\text{diag} = m_N\begin{pmatrix} 0 & 0 & 0 \\ 0 & 1-\frac{\mu}{2} & 0 
    \\ 
    0 & 0 & 1+\frac{\mu}{2} \end{pmatrix}
    ,
    \qquad Y 
    = y 
    \begin{pmatrix} 0& 0 & 0 
    \\
    0 & 
    1+\epsilon_3 & 
    - i(1+\epsilon_2) + \epsilon_1
    \\ 
    0 & 
    i(1+\epsilon_2) + \epsilon_1 & 
    1 - \epsilon_3   
    \end{pmatrix}
\end{equation*}
where $|\epsilon_i|$, $|\mu| \ll 1$. The parameter $\mu$ controls the relative mass splitting between the two species $\mu = (m_{N_3}-m_{N_2})/m_N$ and can be either positive of negative.

In the linear order in terms of the small parameters, Eq.~\eqref{eq:full_seesaw} takes the form 
\begin{equation}
\label{eq:typeIseesaw}
    U^*_{\text{PMNS}} m_\nu^\text{diag} U^\dagger_{\text{PMNS}} = -V_R X V_R^T
\end{equation}
with
\begin{equation}
\label{eq:Xmatrix}
    X = \frac{y^2 v^2}{m_N}\begin{pmatrix}
        0 & 0_{1\times 2} \\  0_{2\times 1} & X_{2 \times 2}
    \end{pmatrix}, \qquad  X_{2\times 2} = -2\epsilon_2 + (- 2 i \epsilon_1 + 2\epsilon_3 + \mu)(\sigma_3 + i \sigma_1)
\end{equation}
where $\sigma_i$ are Pauli matrices.

We adopt the definition in which $m^\text{diag}_\nu =\text{diag}(0, m_2, m_3)$ is the mass-ordered diagonal matrix of active neutrinos with masses:
\begin{align}
    &m_2 = \sqrt{\Delta m^2_\text{21}}\approx \unit[9]{meV}&& \quad m_3 = \sqrt{\Delta m^2_\text{3l}}\approx \unit[50]{meV},&&\quad \text{(NH)}
    \nonumber
    \\
    \label{eq:m2m3masses}
    &m_2 = \sqrt{|\Delta m^2_\text{3l}|}\approx \unit[50]{meV}&& \quad m_3 = \sqrt{|\Delta m^2_\text{3l}|+\Delta m^2_\text{21}}\approx \unit[51]{meV},&&\quad \text{(IH)} 
\end{align}
Moreover, for convenience, we define
\begin{equation}
    \label{eq:tilde-pmns}
    U_{\text{PMNS}} = \tilde U_{\text{PMNS}}  P \begin{pmatrix}
        1 & 0 & 0 \\ 0 & e^{i\eta} & 0 \\ 0 & 0 & 1
    \end{pmatrix}, \qquad P = \begin{pmatrix} 0 & 1 & 0 \\ 0 & 0 & 1 \\ 1 & 0 & 0 \end{pmatrix} \text{ for IH;}\quad P =  1_{3\times 3} \text{ for NH}
\end{equation}
where $\tilde U_\text{PMNS}$ is an experimentally measured matrix as given explicitly in~\cite{Gonzalez-Garcia:2012hef}. In this parameterization, $\eta$ is the single Majorana phase associated with $m_2$ and the matrix $P$ reorders the massive neutrino states to the standard convention.

The perturbation matrix $X$ can be rewritten as 
\begin{equation*}
        X = O^\dagger \begin{pmatrix}
        0 & 0 & 0 \\
        0 & m_2 & 0 \\
        0 & 0 & m_3
    \end{pmatrix} O^*
\end{equation*}
where the explicit form of the rotational matrix $O$ is given by
\begin{multline}
\label{eq:Omatrix}
    O = \frac{1}{\sqrt{2(m_2+m_3)}} \times \\ \times \begin{pmatrix}
        \sqrt{2(m_2+m_3)} & 0 & 0 
        \\
        0 & 
        -i(\sqrt{m_3} e^{-i\beta} \pm \sqrt{m_2} e^{i\beta} ) & 
        \sqrt{m_3} e^{-i\beta} \mp \sqrt{m_2} e^{i\beta} 
        \\
        0 & 
        -(\sqrt{m_3}e^{i\beta} \mp \sqrt{m_2} e^{-i\beta}) & 
        i(\sqrt{m_3}e^{i\beta} \pm \sqrt{m_2} e^{-i\beta})
    \end{pmatrix}  
\end{multline}
and the initial perturbation parameters satisfy the relations
\begin{align}
    \epsilon_1 m_s &= \frac{m_3 - m_2}{4} \sin 2\beta \nonumber\\
    \label{eq:typeIsolution}
    \epsilon_2 m_s &= \pm \frac{\sqrt{m_2 m_3}}{2} \\
    \left(\epsilon_3  + \frac{\mu}{2}\right) m_s &= \frac{m_3 - m_2}{4} \cos 2\beta     \nonumber
\end{align}
where $m_2$, $m_3$ - two active neutrino masses, $\beta$ is a free parameter, and $m_s = \frac{y^2 v^2}{m_N}$ is the neutrino mass scale which one would expect without the approximate lepton symmetry. The last constraint implies that the mass splitting $\mu$ remains a free parameter. Apart from these parameters, the HNLs are parameterized by two sets of couplings
\begin{align*}
    & V^R = i U^*_\text{PMNS} O, && \qquad U^R_{I} = \frac{m_{W_L}^2}{m_{W_R}^2}\\
    & V^L_{\alpha 2} = - i V^L_{\alpha 3} = \frac{e^{i \beta}}{\sqrt{m_2+m_3}}  \tilde U^*_{\text{PMNS}} P \times \begin{pmatrix} 0 \\
    \mp e^{-i\eta}\sqrt{m_2}\\ \sqrt{m_3} \end{pmatrix}, &&\qquad U^L_{2,3} = \frac{\sqrt{2} yv}{m_N}
\end{align*}
which are functions of two free parameters: $\eta$ and $\beta$. The sign change can be compensated for by an appropriate redefinition of the phases $\eta\to \eta +\pi$, $\beta\to \beta+\pi/2$, which would result in an overall phase of $V^R$, $V^L$. Hereafter, we choose the plus sign, whenever the phase redefinition does not affect the conclusions. The model is fully parametrized with 4 general parameters $m_{\text{DM}}$, $m_N$, $U_R$, $U_L$, two internal parameters $\eta$, $\beta$ which fix the flavor matrices $V^L$, $V^R$, and the mass splitting parameter $\mu = \Delta m_N/m_N$.

The approach and results presented here may seem contradictory to the conclusions of~\cite{Nemevsek:2012iq} that the Dirac masses cannot exceed the seesaw limit. The conclusion was based on a naive application of the square root operation to the Dirac mass matrix $M_D \sim \sqrt{m_\nu/M_N}$. It is known that matrix square root can be singular, see App.~\ref{app:square2x2} for a concrete example. In the recent work~\cite{Kriewald:2024cgr}, closed analytic expressions for the square root of $3\times 3$ matrix have been derived showing the same potential singularity.  
The situation described here requires neutrino mass matrix of a specific form, which may be considered as fine-tuning in the bottom-up approach when one reconstructs the properties of heavy neutrinos. In our top-bottom approach, however, this is a perfectly reasonable assumption.    

It is worth to mention several relations in the derived formulae. First, the parameter $\beta$ enters $V^L$ only as an overall phase and does not affect the ratios between the LH couplings. The obtained results coincide with the Casas-Ibarra parameterization in the approximate symmetry limit.

Second, the RH couplings of the DM candidate $V^R_{\alpha 1}$ do not depend on the internal parameters, and their magnitudes are \begin{equation}
\label{eq:VR1}
    |V^R_{\alpha 1}|^2 = |\tilde U^\text{PMNS}_{\alpha 1}|^2 =
    \begin{cases} \begin{pmatrix}
        0.68^{+0.03}_{-0.04}, & 0.07^{+0.18}_{-0.02}, & 0.24^{+0.03}_{-0.18} 
    \end{pmatrix}, &\quad \text{NH} \\
    \begin{pmatrix}
        0.022^{+0.002}_{-0.002}, & 0.57^{+0.04}_{-0.17}, & 0.41^{+0.17}_{-0.04}
    \end{pmatrix}, &\quad\text{IH}
    \end{cases} 
\end{equation}
within the $3\sigma$ confidence intervals~\cite{Esteban:2020cvm} (excluding the Super-Kamiokande data).

The magnitudes of $V^R_{\alpha 2}$, $V^R_{\alpha 3}$ depend on both $\beta$ and $\eta$. However, thanks to the unitarity of the matrix $V^R$, the following combination is fixed: 
\begin{multline}
\label{eq:VR23}
    \frac{|V^R_{\alpha 2}|^2 + |V^R_{\alpha 3}|^2}{2} =\frac{1 - |V^R_{\alpha 1}|^2}{2} \equiv
    \\
    \equiv \langle V^{R}_{\alpha}\rangle^2 = 
    \begin{cases} 
    \begin{pmatrix}
        0.16^{+0.02}_{-0.02}, & 0.47^{+0.01}_{-0.09}, & 0.38^{+0.09}_{-0.02}
    \end{pmatrix}, & \text{NH} \\
    \begin{pmatrix}
        0.489^{+0.001}_{-0.001}, & 0.22^{+0.09}_{-0.02}, & 0.30^{+0.02}_{-0.09}
    \end{pmatrix}, &\text{IH}
    \end{cases} 
\end{multline}

\begin{figure}[t!]
    \centering
    \includegraphics[height = 0.42\textwidth]{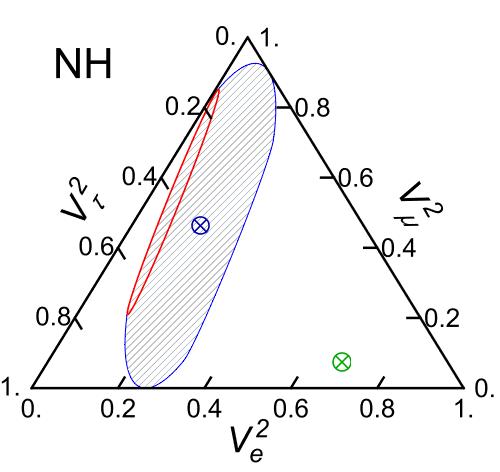}~
    \includegraphics[height = 0.42\textwidth]{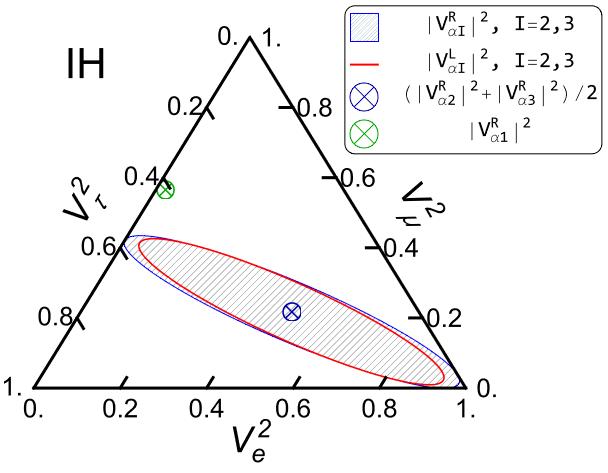}
    \caption{Ternary plots for the couplings $|V^L|^2$, $|V^R|^2$ in the type-I seesaw with approximate lepton symmetry, for the normal (left) and inverse (right) neutrino hierarchies. The red lines show the range of possible values of the LH couplings $|V^L_{\alpha 2}|^2 = |V^L_{\alpha 3}|^2$. The green point shows a fixed value of the RH couplings $|V^R_{\alpha 1}|$ of the DM candidate, given by Eq.~\eqref{eq:VR1}. The dashed blue region represents the range of the values for $|V^R_{\alpha 2}|^2$, $|V^R_{\alpha 3}|^2$. The corresponding pairs of points are located symmetrically around the blue circle, depicting the fixed value $\langle V^R_\alpha\rangle^2 \equiv (|V^R_{\alpha 2}|^2+|V^R_{\alpha 3}|^2)/2$, see Eq.~\eqref{eq:VR23}. The plot uses the best-fit neutrino oscillation parameters~\cite{Esteban:2020cvm} (without the SK data) and does not include the corresponding experimental uncertainties.}
    \label{fig:ternary}
\end{figure}

The summarizing ternary plots of the values of $V_L$, $V_R$ are shown in Fig.~\ref{fig:ternary}.

\subsection{Mixed type-I and II seesaw}
\label{sec:typeIIseesaw}

With the type-II contribution, the seesaw equation becomes
\begin{equation}
    \label{eq:typeI+IIequation}
    U_\text{PMNS}^* m^\text{diag}_\nu U_\text{PMNS}^\dagger =  - V_R X
    V_R^T + \kappa (m_2 + m_3) V_R^* \begin{pmatrix} 0 & 0 & 0 \\ 0 & 1 & 0 \\ 0 & 0 & 1 \end{pmatrix} V_R^\dagger
\end{equation}
where the introduced a dimensionless parameter 
\begin{equation}
    \kappa = \frac{v_L m_N}{v_R(m_2 + m_3)}
\end{equation}
The parameters $m_2$, $m_3$ keep their original definition as given in Eq.~\eqref{eq:m2m3masses}.

This adjustment leads to a few qualitative changes: first, the mass of the lightest neutrino in $m_\nu^\text{diag}$ no longer has to be equal zero, despite the fact that both terms in the r.h.s. only have rank two. The reason is that, in general, these two terms are not simultaneously diagonalizable, and one can easily check that a sum of such two matrices yields a rank three matrix. Second, the neutrino mass matrix must be specified with three Majorana phases, rather than only one as in Eq.~\eqref{eq:tilde-pmns}. One additional phase is obviously attributed to the new neutrino mass eigenvalue. The appearance of the second phase can be explained in the following way: the previous type-I seesaw relation~\eqref{eq:typeIseesaw} has a phase rotation symmetry $U_\text{PMNS} \to e^{i\phi} U_\text{PMNS}$, $V_R \to e^{-i\phi} V_R$ that has been ignored by choosing a specific form of $U_\text{PMNS}$. With two terms involving both $V_R$ and its complex conjugated $V_R^*$, this symmetry does not hold, and a change in the overall phase in $U_{\text{PMNS}}$ does modify the magnitudes of the entries of the matrix $V_R$.\footnote{One can compare this to the equation of the form $z^2 = x^2 - a (x^*)^2$. For $b=0$, phase rotation in $z$ can be trivially translated into the change in $x$, but this is no longer the case for $a\neq 0$.}

This is a fourth-order equation in the desired $V_R$ matrix, which complicates the analytic solution. Therefore, we solve the equation numerically in the following form:
\begin{equation*}
    V_R = i U_\text{PMNS} e^{iC} O(\beta)
\end{equation*}
where the matrix $O$ is given by~\eqref{eq:Omatrix} and $C$ is a hermitian correction matrix, being zero for $\kappa = 0$. Multiplication $V_R \to V_R \cdot \text{diag}(e^{i\phi}, 1, 1)$ by a phase leaves the equation invariant, meaning that one free parameter remains. We add the constraint $C_{11} = 0$ to fix the solution uniquely.

The type-I mass matrix $X$ has three independent parameters $\epsilon_1$, $\epsilon_2$, $\epsilon_3 + \mu/2$. To use the same convenient parametrization~\eqref{eq:typeIsolution}, we introduce two adjustment parameters $\delta \mu_1$, $\delta \mu_2$: 
\begin{align*}
    \epsilon_1 m_s &= \left(\frac{m_3 - m_2}{4} + \delta \mu_1(m_2+m_3)\right) \sin 2\beta
    \\
    \epsilon_2 m_s &= \frac{\sqrt{m_3 m_2}}{2} + \delta \mu_2(m_2+m_3),
    \\
    \left(\epsilon_3+\frac{\mu}{2}\right)m_s &= \left(\frac{m_3 - m_2}{4} + \delta \mu_1(m_2+m_3)\right)\cos 2\beta
\end{align*}
With this choice or parametrization, the $\beta$ parameter drops out from equation~\eqref{eq:typeI+IIequation} identically. 

In summary, the matrix equation~\eqref{eq:typeI+IIequation} has 12 real equations (taking into account that the l.h.s. matrix is symmetric) with eight entries of the hermitian matrix $C$, two parameters $\delta \mu_1$, $\delta \mu_2$, and four additional parameters: mass of the lightest neutrino, its associated phase, and two other phases entering a modified version of Eq.~\eqref{eq:tilde-pmns}:
\begin{equation*}
    U_{\text{PMNS}} = \tilde U_{\text{PMNS}}  P \begin{pmatrix}
        1 & 0 & 0 \\ 0 & e^{i\eta_1} & 0 \\ 0 & 0 & e^{i\eta_2}
    \end{pmatrix}
\end{equation*}
To match the number of equations to the number of free parameters, we treat $\eta_1$, $\eta_2$ as fixed parameters, and search for the solution in the remaining twelve parameters. We build the solution iteratively: starting from $\kappa = 0$ and the exact solution, we increase $\kappa$ by a small step and find the values of the unknown parameters by minimizing the total squared error between the entries on the left and right sides. Once found, $\kappa$ is increased once again and the process is repeated, producing a unique one-parametric solution for the matrix $C(\kappa)$ and hence $V_R$ for each fixed $\eta_1$, $\eta_2$. 

The strictest bounds on the total neutrino mass come from cosmological observations. We adopt the Planck limit $\sum_\nu m_\nu < \unit[0.12]{eV}$~\cite{Planck:2018vyg}, although this constraint is tentative and could actually be relaxed~\cite{DiValentino:2021imh}.
Our numerical solutions that violate the bound are excluded.  Notably, for most of the combinations of $\eta_1$, $\eta_2$, the minimal neutrino mass stays at a level below the bound even for the largest $\kappa$ we considered. 

\begin{figure}[t!]
    \centering

    \includegraphics[width = 0.4\textwidth]{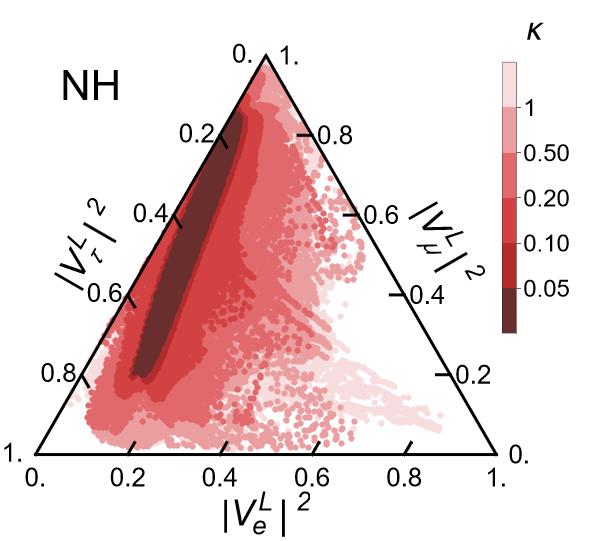}~
    \includegraphics[width = 0.4\textwidth]{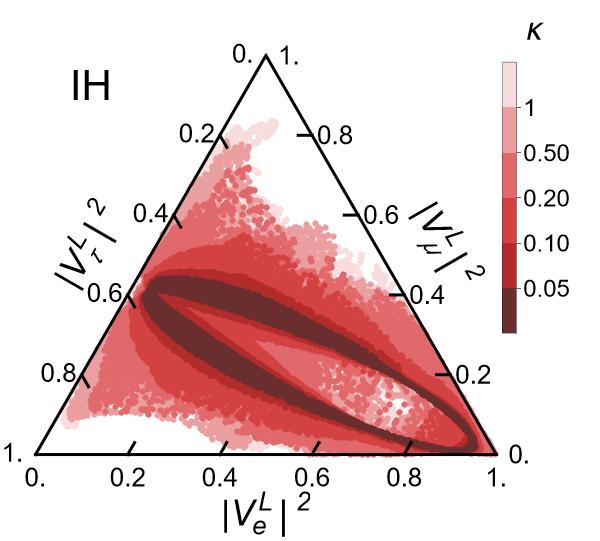}

    \includegraphics[width = 0.4\textwidth]{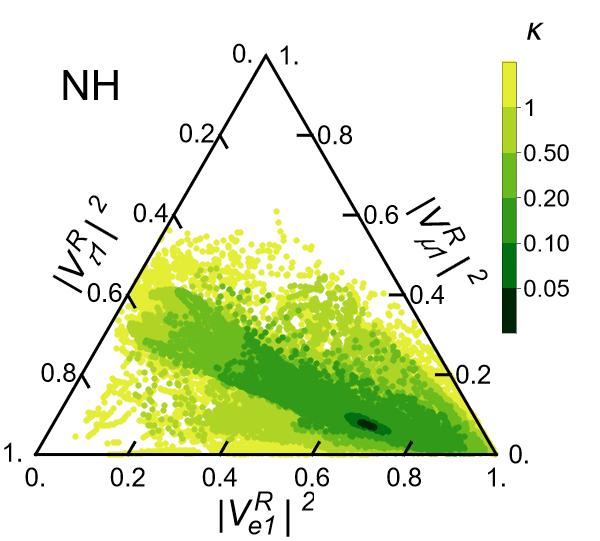}~   
    \includegraphics[width = 0.4\textwidth]{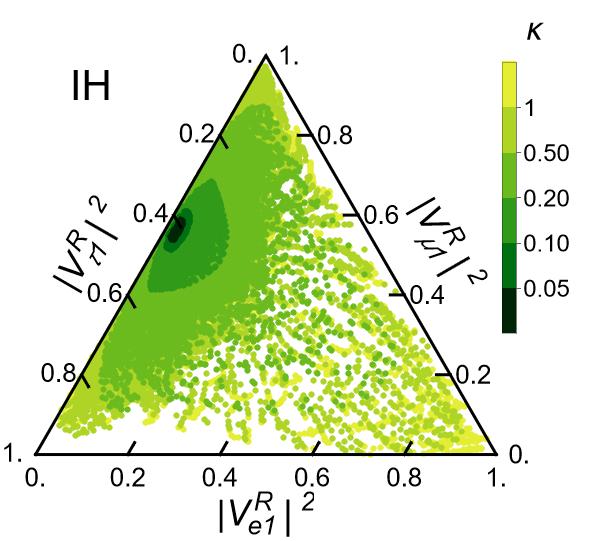}
    
    \includegraphics[width = 0.4\textwidth]{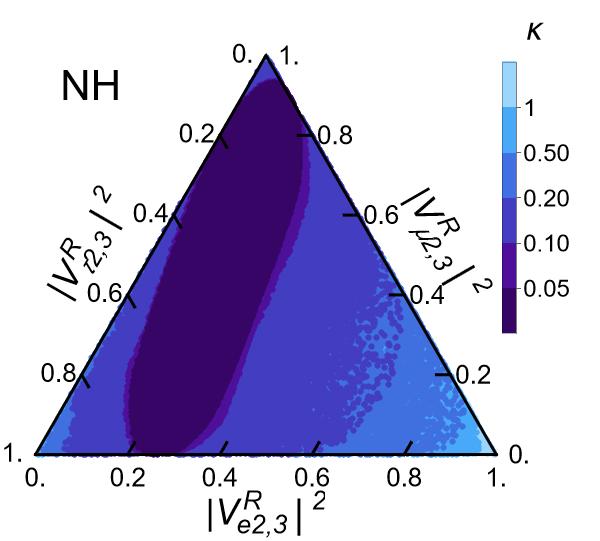}~ 
    \includegraphics[width = 0.4\textwidth]{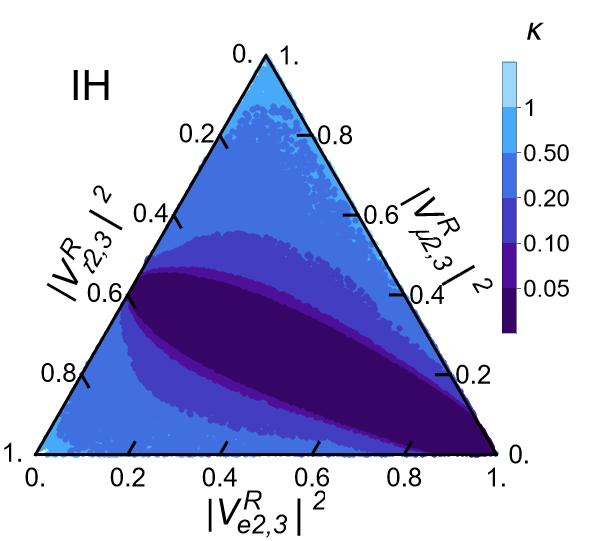}

    \caption{Numerical scan for the possible values of $|V^L_{\alpha2}|^2=|V^L_{\alpha3}|^2$ (top), $|V^R_{\alpha1}|^2$ (middle), and $|V^R_{\alpha2, 3}|^2$ (bottom) for the included type-II seesaw contribution, controlled by the $\kappa = v_L m_N/v_R (m_2+m_3)$ parameter. The right-handed couplings averaged over the heavy species $\langle V^R_\alpha \rangle^2 = (|V_{\alpha 2}^R|^2+|V_{\alpha 3}^R|^2)/2$ can be obtained from the middle plot, using the unitarity of the coupling matrix. The left (right) column corresponds to the normal (inverse) hierarchy of neutrino masses. The colors match the legend of Fig.~\ref{fig:ternary}. The scan is performed in the parameter space of $\kappa \in (0,20)$ and two Majorana phases $\eta_1$, $\eta_2$. The numerical results indicate that for $\kappa<0.2$ the corrections to $|V^{R,L}|^2$ are limited to $\lesssim0.1$. Higher values of $\kappa$ effectively allow for arbitrary couplings but also imply some amount of fine-tuning, to ensure cancellation between the type-I and type-II contributions.}
    \label{fig:kappa_correction}
\end{figure}

We performed the scan in the range $\kappa \in (0,20)$ and, for each step, found the minimum of the total squared error with the tolerance of $10^{-8}$. The possible values of the left and right-handed couplings are shown in Fig.~\ref{fig:kappa_correction}.

\subsection{Type-I with $b\neq 0$}
\label{sec:typeInonzerob}

To this point, we neglected the contribution of the $b$ angle in the $\Phi$ vacuum expectation value. The two changes the angle introduces are a correction to the seesaw relation and mixing between the charged bosons, leading to RH interactions of HNLs through the Standard Model $W$-boson. We first argue that the seesaw contribution naturally requires $b$ to be small. For $b\ll1$, the neutrino Dirac mass matrix is modified to
\begin{equation*}
    v Y_\nu \to v Y_\nu - m^\text{diag}_l b e^{ia} 
\end{equation*}
The second term will result in the contribution to the neutrino mass matrix at the scale of $m_\nu \sim b^2 m_\tau^2/m_N$
where the charge lepton mass matrix is dominated by the tau-lepton mass. This relation can be used to put a crude constraint 
\begin{equation}
\label{eq:b_crude_limit}
    b \lesssim 10^{-5} \sqrt{\frac{m_N}{\unit{GeV}}}
\end{equation}
which are much stricter for HNLs below the TeV scale than the current limits from the amount of CP violation in the quark sector, $b \sin a \lesssim m_b/m_t$~\cite{Maiezza:2010ic,Senjanovic:2014pva, Senjanovic:2015yea}. The effective coupling for the $W$-mediated coupling has the order of~\cite{Senjanovic:1978ev}:
\begin{equation}
    \xi_W \sim \frac{m_W^2}{m^2_{W_R}} \,b = U_R \, b
\end{equation}
which is further suppressed compared to the effective coupling $U_R$ of the $W_R$-mediated interaction.

The constraint~\eqref{eq:b_crude_limit} is crude because one can hypothesize an adjustment of the $Y_\nu$ matrix to cancel the contribution of the charge lepton mass matrix. While one cannot exclude such a possibility, this adjustment faces two problems. First, a change in a hermitian matrix cannot cancel the CP-violating part, proportional to $b \sin a$. Second, the misalignment between the heavy states in the charged lepton and sterile neutrino sectors implies that such an adjustment would necessarily include a contribution of the lightest $N_1$ neutrino species, which we forbid in the considered model. In any case, values of $b$ larger than given by the constraint do require some amount of fine-tuning on top of the assumed approximate lepton symmetry of the HNL pair.

In this section, we assume that $b$ is small to be treated as a correction and repeat the same procedure as described in Sec.~\ref{sec:TypeIcase}. The linearized equation reads:
\begin{multline}
    U_\text{PMNS}^* m^\text{diag}_\nu U_\text{PMNS}^\dagger =  - V_R X
    V_R^T + \frac{\tilde b e^{ia}}{\sqrt{2}} (m_2 + m_3)
    \\
    \times \left[
        \begin{pmatrix}
            0 & 0 & 0 \\
            0 & m_\mu/m_\tau & 0 \\
            0 & 0 & 1
        \end{pmatrix} V_R   \begin{pmatrix}
            0 & 0 & 0 \\
            0 & 1 & -i \\
            0 & i & 1
        \end{pmatrix} V_R^T + (\text{transpose})
    \right]
\end{multline}
where we have kept muon mass $m_\mu$ but neglected electron mass. The dimensionless parameter, equivalent to $\kappa$ in the case of type-II correction, is given by
\begin{equation}
\label{eq:tilde_b}
    \tilde b = b \frac{\sqrt{2} y v m_\tau}{m_2+m_3} = b \frac{m_\tau U}{m_2 + m_3}, \qquad \tilde b \ll \sqrt{\frac{m_N U^2}{m_2 + m_3}}
\end{equation}
The inequality represents the necessary condition for the validity of the linear expansion, which simply requires that the quadratic in $b$ term is negligible. Namely, $b$ must be sufficiently below the constraint Eq.~\eqref{eq:b_crude_limit}.

\begin{figure}[t!]
    \centering

    \includegraphics[width = 0.4\textwidth]{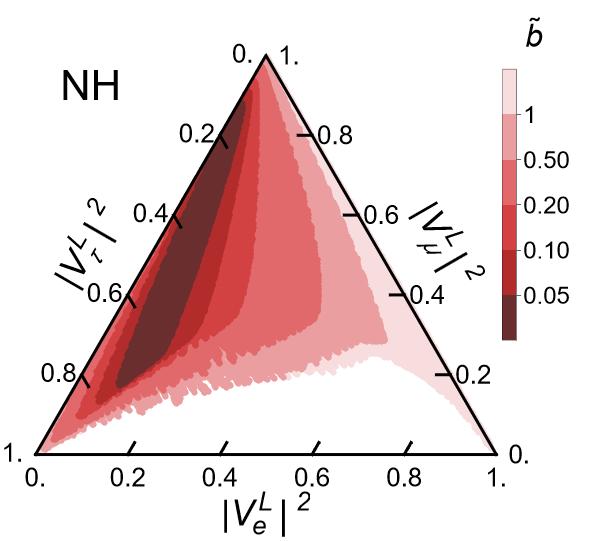}~
    \includegraphics[width = 0.4\textwidth]{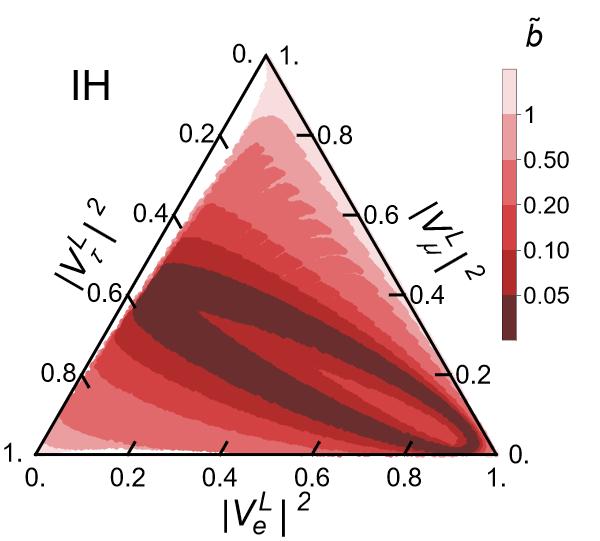}
    
    \includegraphics[width = 0.4\textwidth]{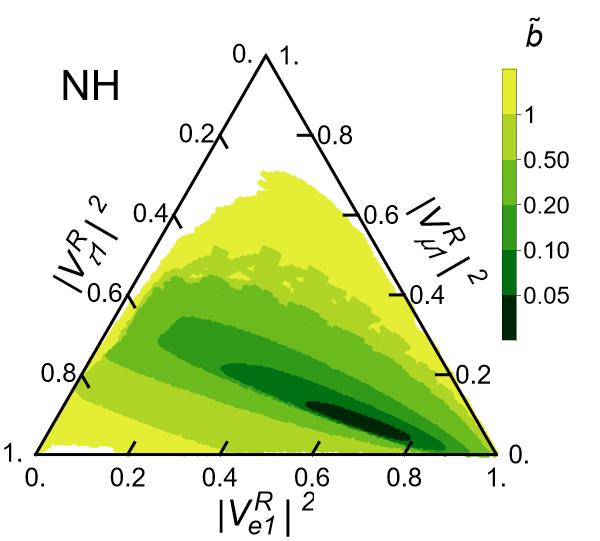}~   
    \includegraphics[width = 0.4\textwidth]{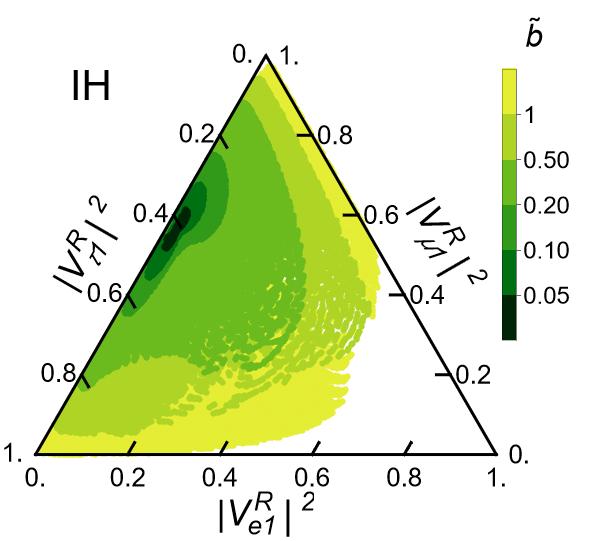}  
    
    \includegraphics[width = 0.4\textwidth]{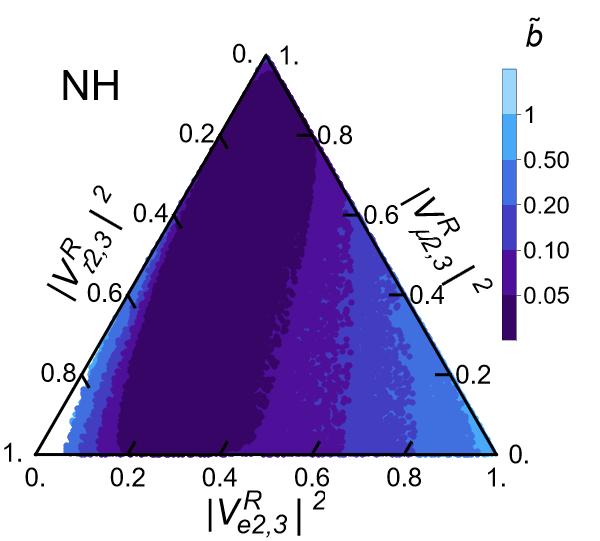}~
    \includegraphics[width = 0.4\textwidth]{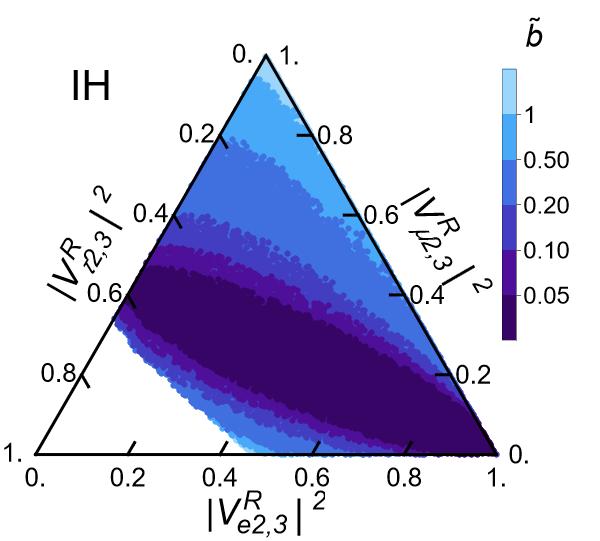}
    
    \caption{Numerical scan for the possible values of $|V^L_{\alpha2}|^2 = |V^L_{\alpha3}|^2$ (top), $|V^R_{\alpha1}|^2$ (middle), and $|V^R_{\alpha2, 3}|^2$ (bottom) for various $\tilde b$, defined in Eq.~\eqref{eq:tilde_b}. The left (right) column corresponds to the normal (inverse) hierarchy of neutrino masses.}
    \label{fig:CP_correction}
\end{figure}

We employ the same parametrization as for the type-II correction. The sole difference is the drop of the redundant Majorana phase $\eta_2$, which in this case can actually be absorbed with a trivial phase rotation of both $U_\text{PMNS}$ and $V_R$. The range of the numerical scan is limited to $\tilde b<10$. The plots with the possible values of $|V^R|^2$, $|V^L|^2$ are given in Fig.~\ref{fig:CP_correction}.

\section{Experimental probes}

In this section, we analyze various experimental probes which may be used to identify the model. For simplicity, we \textbf{consider only the minimal case of Sec.}~\ref{sec:TypeIcase}, assuming that the corrections due to the type-II seesaw and nonzero $b$ are negligible. Possible non-minimality of the model, as well as experimental details related to the precision of such probes, can and do affect the conclusions of this session. We deliberately omit these details for the sake of generality.

\subsection{Sensitivity of FCC-ee in the $Z$-pole mode}

\label{sec:FCCee}

The best sensitivity to minimal heavy neutral leptons in the $\sim \unit[10-100]{GeV}$ mass range is offered by the FCC-ee, operating at the $Z$-pole mass~\cite{Blondel:2014bra,Blondel:2022qqo}, thanks to the abundant sample reaching more than $10^{12}$ $Z$-bosons. The production in the decay of $Z$-boson occurs through either the $Z\to N \nu$ process, mediated by the left-handed interaction, or the yet-unaccounted mixing between $Z$ and a possible heavy state $Z_R$, which would enable the $Z\to N N$ decay. Our analysis focuses on the properties of fermions and has been largely independent of the details of the bosonic sector of the model, only relying on the existence of $W_R$. The process $Z\to N N$, however, does require model-dependent details. Let us, for a moment, ignore this production channel. The decays of the HNLs can still be mediated by both left and right-handed interaction, thus offering a way to probe nonminimal HNL interactions. Additional interactions may enhance the signal rate to the observable level in parts of the parameter space otherwise unreachable. This refers specifically to the light HNLs with small coupling constant $U^2_L$, for which, in LH-interaction case only, the decay length reaches macroscopic values and suppresses the probability of the decay within the detector system. With the new interaction, the HNL decay width is increased, forcing the HNL to decay faster and avoiding the mentioned obstacle. 

To estimate the sensitivity of an experiment, hosted by FCC-ee, let us provide simple scaling arguments. At the lower bound of sensitivity~\cite{Bondarenko:2019yob}, the number of events is proportional to:
\begin{equation}
\label{eq:nev_FCC}
    N_\text{ev} \propto \underbrace{\left[U^2_L \text{Br}_{Z\to \nu N} + c_Z U_R^2 \text{Br}_{Z\to NN}\right]}_{\text{production}} \quad\times \quad \underbrace{[U_L^2 \cdot \text{Br}_{\text{vis.},L} + U_R^2 \cdot \text{Br}_{\text{vis.},R} ]}_{\text{decay width}}
\end{equation}
where the production branchings 
\begin{align*}
    \text{Br}_{Z\to \nu N} & =  \left(1 - \frac{m_N^2}{m_Z^2}\right)^2\left(1 + \frac{m_N^2}{2 m_Z^2}\right) \\
    \text{Br}_{Z\to N N}  & =  \left(1 - \frac{m_N^2}{m_Z^2}\right) \sqrt{1 - \frac{4 m_N^2}{m_Z^2}}
\end{align*}
carry the kinematic dependence on the HNL mass,
while $\text{Br}_{\text{vis}, L(R)}$ stands for the visible branching ratio for the pure left(right)-handed interactions. The parameter $c_Z$ contains the $ZNN$ couplings, except for the overall suppression scale $U^2_R$, and in the minimal LRSM is given by~\cite{Duka:1999uc}:
\begin{equation*}
        c_Z = \frac{1}{4} \frac{\cos^4 2\theta_W}{\cos^4 \theta_W} \approx 0.04
\end{equation*}
This parameter is notably small. If the production is dominated by the $Z\to NN$ decay, the decay width is also dominated by $U^2_R$ and the LH interaction can be neglected. If $U_R\lesssim U_L$, both production and decay are dominated by LH interactions. Nevertheless, there exists a region in the parameter space between:
\begin{equation*}
    \frac{U^2_L}{c_Z} \frac{\text{Br}_{Z\to \nu N}}{\text{Br}_{Z\to NN}} \gtrsim U^2_R \gtrsim U^L_2 
\end{equation*}
in which production and decay become parametrically independent.

For a more rigorous analysis of the FCC-ee sensitivity to heavy neutral leptons, we use the setup described in~\cite{Blondel:2022qqo}, with the expected  $5\cdot 10^{12}$ $Z$-bosons. Specifically, we assume the CLD design, with the geometry of a cylinder with length $\unit[8.6]{m}$ and radius $\unit[5]{m}$. The decay length of the HNL should exceed $\unit[400]{\mu m}$. The number of required decay events is four.

For the sake of generality, we neglect masses of the charged leptons and $u,d,s,c$ quarks, as neglect the decay of HNLs into a $b$-quark. The last assumptions can safely done for $N\to l q_u b$ decays thanks to the smallness of $ub$, $sb$ entries of CKM matrix. For the neutral current decays $N\to \nu bb$, however, this can lead to an error of the order at most $5\%$ level in the decay width. Finally, we assume that cross-HNL decays $N_I \to N_J l_\alpha l_\beta$ are absent and only one HNL species is produced. These assumptions, having a minor effect on the results, allow us to eliminate the flavor-specific details and build general bounds which valid for any combinations of $V^{L/R}_{\alpha I}$.

\begin{figure}[t!]
    \centering
    \includegraphics[height = 0.4\linewidth]{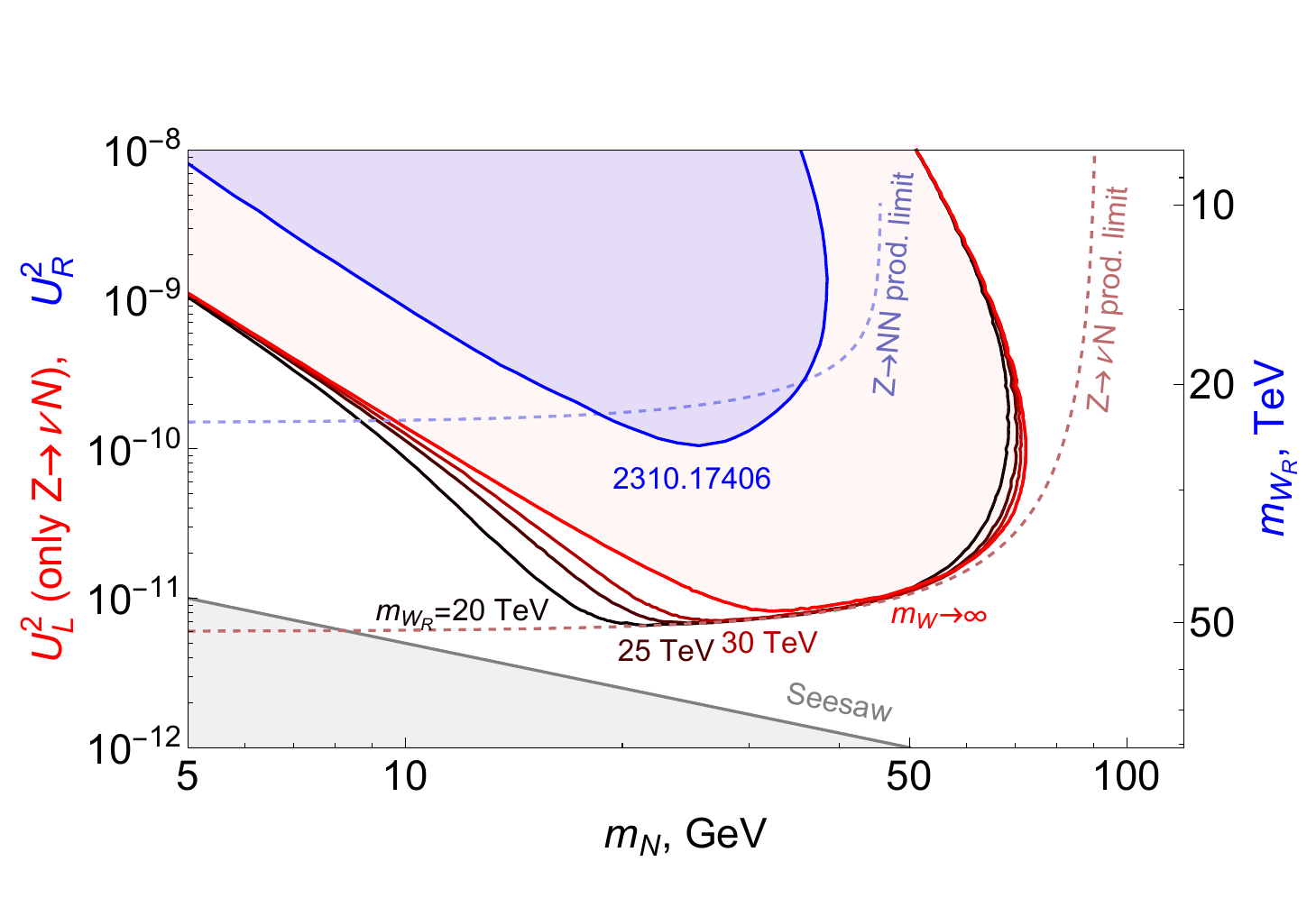}~
    \includegraphics[height = 0.4\linewidth]{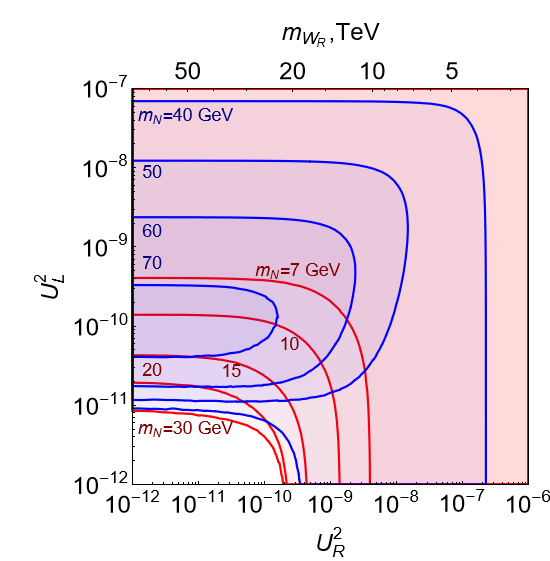}
    
    \caption{\textit{Left:} sensitivity contours for $U^2_L$ (red-shaded lines), assuming HNL production \textbf{only} in $Z \to \nu N$ decays. The filled red region is the sensitivity of the standard model with only left-handed interactions. Below the light-red line denoted as \textit{Z prod limit} in the $U^2_L$ parameter space, the number of HNLs produced is below the required limit of four particles. With $W_R$ masses around $20-\unit[40]{TeV}$, the sensitivity to small $U_L^2$ in the HNL mass range $\unit[20-30]{GeV}$ improves at the lower bound. For larger $W_R$ mass, RH starts to dominate, as demonstrated by the proper sensitivity reach to $U_R^2$ (blue line) from~\cite{Urquia-Calderon:2023dkf} (one displaced vertex, PMNS mixing), which includes all production channels. The light-blue line corresponds to the production limit from $Z\to NN$ in the $U^2_R$ space. \textit{Right:} estimated sensitivity contours in the $(U^2_L, U^2_R)$ space for various choices of the HNL mass, from $\unit[7]{GeV}$ to $\unit[70]{GeV}$, including \textit{both} $Z\to \nu N$ and $Z\to NN$ decay channels with $c_Z = 0.04$. }
    \label{fig:FCCee}
\end{figure}

Under these assumptions, the full decay width can be approximated as
\begin{equation*}
    \Gamma_{L/R} = 2 U^2_{L,R}\times C_{L/R} \times \frac{G_F^2 m_N^5}{192\pi^3},
\end{equation*}
where prefactor 2 counts the charge-conjugated decay modes for a Majorana fermion. The coefficients are
\begin{equation*}
        C_R = 6, \qquad
        C_L = \frac{1}{12}(153 - 84 \sin^2 \theta_W + 152 \sin^4 \theta_W) =11.8
\end{equation*}
The decay width of the only invisible mode is $\Gamma(N\to 3 \nu) = 2 U^2_L \gamma_0$. The final note is the account of different HNL momentum for the two production modes, which determine the decay length.

The sensitivity of FCC-ee with to HNLs is shown in Fig.~\ref{fig:FCCee}. For smaller mixing angles and masses below $\unit[30]{GeV}$, the LH sensitivity drops due to the large lifetime of HNLs. The RH interactions mediated by $W_R$ with mass $\unit[20-40]{GeV}$ improve the sensitivity to $U_L$ in this region by forcing the particles to decay more promptly. For small $U^2_L$ and masses above $\unit[30]{GeV}$, the particle lifetime is sufficiently small, and the sensitivity is saturated by the production limit: the LH interaction is too tiny to produce any HNLs below the bound.

\subsection{(De)coherence of the quasi-Dirac pair}

Observation of lepton number violating (LNV) processes in collider experiments would be a smoking gun for new physics, as illustrated by the famous Keung-Senjanovic process~\cite{Keung:1983uu} for searching for the LRSM right-handed neutrinos. Lepton number violation is closely related to leptogenesis and the nature of active neutrino masses, which, if Majorana, should exhibit themselves in $0\nu\beta\beta$ decays~\cite{Vergados:2012xy}. In the scenario of quasi-Dirac HNLs, LNV happens due to oscillations between the lepton and antilepton-like parts of the pair. The analysis of the properties of these oscillations and various collider probes can be found in~\cite{Antusch:2017ebe, Antusch:2017pkq, Drewes:2019byd, Schubert:2022lcp}. If one of the two SM leptons, required to establish LNV directly, escapes detection, e.g., in the form of neutrino at lepton colliders~\cite{Antusch:2023jsa, Mikulenko:2023ezx} or is absorbed in the target in case of beam dump experiments~\cite{Tastet:2019nqj}, kinematic information may still be used to probe the nature of HNLs. Although oscillations are believed to be directly defined by the innate properties of the new particles (namely, mass splitting), Ref.~\cite{Antusch:2023nqd} claims that dynamical effects, arising from the uncertainty of particle energy, may lead to decoherence between the two species and spoil the connection between the observed LNV and particle parameters. In what follows, we study the properties of HNL interactions in two limits, assuming that the oscillations are either very slow or very fast and treating this choice as an independent model parameter.

At the moment of HNL creation, the two species are produced in a superposition. If the oscillations between them become too rapid at the moment of decay, the superposition becomes decoherent: the two particles can be treated as independent, and any observable signal may be computed as simply a sum of two separate contributions. If the oscillations are sufficiently slow, the HNLs remain coherent and interfere with each other, affecting the decay pattern of such a superposition. In the pure left-handed case, the final states that violate lepton number conservation cancel each other as a manifestation of the quasi-Dirac nature of the HNL pair. The situation becomes more complicated for the right-handed interactions, as they do not respect the imposed approximate lepton symmetry, which is attributed only to the specific choice of the Dirac mass matrix without any particular symmetry-driven restrictions on $V^R$.

Below, we analyze the coherent and decoherent cases. For simplicity, we assume a hierarchical relation between the small couplings $U^2_L$, $U^2_R$, such that the two interactions do not mix in production/decay. We assume that an HNL is produced together with a lepton of flavor $\alpha$, and interacts with a $\beta$-flavored (anti)lepton.

\begin{enumerate}
    \item[(a)] \textbf{Rapid decoherence.}

If the two HNLs are treated as independent particles, the total number $n_I$ of HNLs created with the $\alpha$ lepton flavor is simply proportional to $|V_{\alpha I}|^2$.
with $V$ standing for either $V^R$ or $V^L$, depending on the type of interactions. Each of the HNLs then decays into $\beta$-flavored leptons and antileptons independently with equal probabilities. 

The number of events with $l_\alpha$ accompanying the HNL production and $l_\beta$ or $\bar l_\beta$ in the final state becomes, up to the overall normalization, is given by the matrix:
\begin{equation}
\label{eq:decoherent_decays}
    n_{\alpha\beta} = n_{\alpha \bar \beta} \propto P_\alpha D_\beta ( |V_{\alpha 2}|^2 |V_{\beta2}|^2+|V_{\alpha3}|^2|V_{\beta 3}|^2 )
\end{equation}
where $P_\alpha$ ($D_\beta$) is the kinematic factor, carrying the dependence on the mass of the charged lepton. The kinematic matrix $P_\alpha D_\beta$ is therefore determined by the particle mass (measured by some other means) and the experimental setup, which may have different acceptance and detection efficiencies for different leptons. In the case of LH interactions, there is no difference between $I=2,3$, resulting in the standard product $|V^L_{\alpha 2}|^2 \cdot  |V^L_{\beta 2}|^2$. In the RH case, the event matrix depends explicitly on both internal model parameters --- phases $\beta$ and $\eta$.

\item[(b)] \textbf{Coherent pair.}

In the opposite scenario, the HNLs form a Dirac fermion. In a sequential process of HNL production and decay, the intermediate wave function is a mixture of $N_2$ and $N_3$. Despite each of them being Majorana particles and having equal probabilities of transforming into either a lepton or an antilepton, the same conclusion does not apply to their superposition because of interference. The combined wave function created after the interaction is determined by the corresponding currents:
\begin{align*}
    J_{L,\alpha}& \propto V^L_{\alpha2}\bar l (N^c_2 - i N^c_3) + \text{h.c.} \\
    J_{R,\alpha} &\propto V^R_{\alpha 2} \bar l N_2 + V^R_{\alpha 3} \bar l N_3 + \text{h.c.} 
\end{align*}

The matrix of the number of events with different lepton flavors gets the following form:
\begin{align*}
    n_{\alpha \bar \beta} &\propto P_\alpha D_\beta |V_{\alpha 2}V^{*}_{\beta 2} + V_{\alpha 3} V^{*}_{\beta 3}|^2 && \text{(LNC)}
    \\
    n_{\alpha \beta} &\propto P_\alpha D_\beta |V_{\alpha 2}V_{\beta 2} + V_{\alpha 3} V_{\beta 3}|^2 && \text{(LNV)}
\end{align*}
For LH interactions $V^L_{\alpha 2} = -i V^L_{\alpha 3}$, the standard result is the absence of a lepton number violation in the final state. For RH interactions, both matrices are always nonzero.

In the case of right-handed production and decay, may be simplified to:
\begin{align}
    n_{\alpha \bar \beta} \propto    \label{eq:coherent_decays_lnc}
    &   P_\alpha D_\beta |\delta_{\alpha \beta} - 
    [U_{\text{PMNS}}]_{\alpha 1} [U_{\text{PMNS}}]^*_{\beta 1}|^2 
    \\
    n_{\alpha \beta} \propto\label{eq:coherent_decays_lnv}
    & \frac{P_\alpha D_\beta}{(m_2+m_3)^2} \left|\left(U^*_\text{PMNS} \left(\begin{smallmatrix} 
    0 & \,\,0 & \,\,0 \\
    0 & \,\,\mp i \sqrt{m_2 m_3} & \,\,m_3 - m_2 \\
    0 & \,\,m_3 - m_2 & \,\,\mp i\sqrt{m_2 m_3}
    \end{smallmatrix}\right) U^\dagger_\text{PMNS}\right)_{\alpha \beta}
    \right|^2
\end{align}
with the PMNS matrix given by~\eqref{eq:tilde-pmns} and the plus-minus sign corresponding to that of Eq.~\eqref{eq:typeIsolution}. The Majorana phase $\eta$ only affects the matrix of LNV events, while the LNC event matrix is fixed. 

If the pair remains coherent, the dependence of the event matrices on the parameter $\beta$ vanishes as a result of the underlying symmetry between the two HNLs.

\end{enumerate}

In some experimental setups, the initial lepton $l_\alpha$ may escape detection. This applies to FCC-ee in the $Z$-pole mode, where the HNL can be produced in $Z\to \nu N$ decays with an unobservable neutrino, and for beam-dump experiments, in which all production-related information gets lost in the target. In this case, the event matrices must be summed up over $\alpha+\bar \alpha$. If the kinematics factors $P_\alpha$ are equal all lepton flavors, most of the event matrices are reduced to the same parametric dependence thanks to the unitarity of $V^R$:
\begin{equation}
\label{eq:reduced_nll}
    \sum_\alpha n_{\alpha \beta} \propto D_\beta (|V^R_{\alpha2}|^2 + |V^R_{\alpha 3}|^2), \qquad \text{ if } P_{\alpha} = \text{const}
\end{equation}
proportional to the averaged couplings $\langle V^R_\alpha\rangle^2$ of Eq.~\eqref{eq:VR23}, which are \textit{uniquely fixed} up to neutrino hierarchy. The list of matrices that reduce to this relation is given in Table.~\ref{tab:reduced_nll}. In other words, all the internal information of the models vanishes, and the apparent experimental signal is equivalent to the one from a single HNL with a very specific set of couplings, \textit{determined only by neutrino mass hierarchy}. This can be used a perfect benchmark model of quasi-Dirac HNLs within the minimal seesaw scenario of Sec.~\ref{sec:TypeIcase}. For the nonminimal seesaw scenarios, the same result~\eqref{eq:reduced_nll} holds with the appropriate values of $(V^R_\alpha)^2$, which are no longer fixed.

\begin{table}[h]
    \centering
    \begin{tabular}{|c|c|c|}
        \hline
         & Decoherent case & Coherent case  \\
         \hline
        $R$ prod., $R$ decay & LNC, LNV  & LNC, LNV \\
        \hline
        $L$ prod., $R$ decay & LNC, LNV & LNV + LNC \\
        \hline
        
    \end{tabular}
    \caption{Event matrices that reduce to the same parametric form of Eq.~\eqref{eq:reduced_nll} after summation over the initial lepton flavor $\alpha$. For LH production and RH decay, the relation only holds for the sum of the lepton number violating and conserving decays; in all other cases, the relation holds for both types separately.}
    \label{tab:reduced_nll}
\end{table}

Let us now discuss the implications of the event matrix expressions for differentiating between the coherent and decoherent cases. First, let us start with the mentioned \textbf{FCC-ee in the $Z$-pole regime}, specifically the $Z\to \nu N$ channel. Production from $ee\to Z\to NN$ will be discussed below. The neutrino in the $Z\to \nu N$ decay cannot be detected, therefore, neither the associated lepton flavor or charge can be identified. One has to rely only on the distribution of different lepton flavors in the final state, compared with Eq.~\ref{eq:reduced_nll}. On the one hand, a deviation from this strict prediction would immediately exclude the considered minimal model. On the other hand, if the experimental results do agree with the model, it becomes significantly more challenging to establish the nature of the new fermions. A possible solution is to employ the nonzero polarisation of the $Z$-bosons and the resulting features in the kinematic distribution of the final lepton~\cite{Blondel:2021mss}. Another possibility consists of observing oscillations directly as time-dependent variance in the properties of the HNLs~\cite{Antusch:2023jsa}.

\begin{figure}[t]
    \centering
    \includegraphics[height = 0.4\textwidth]{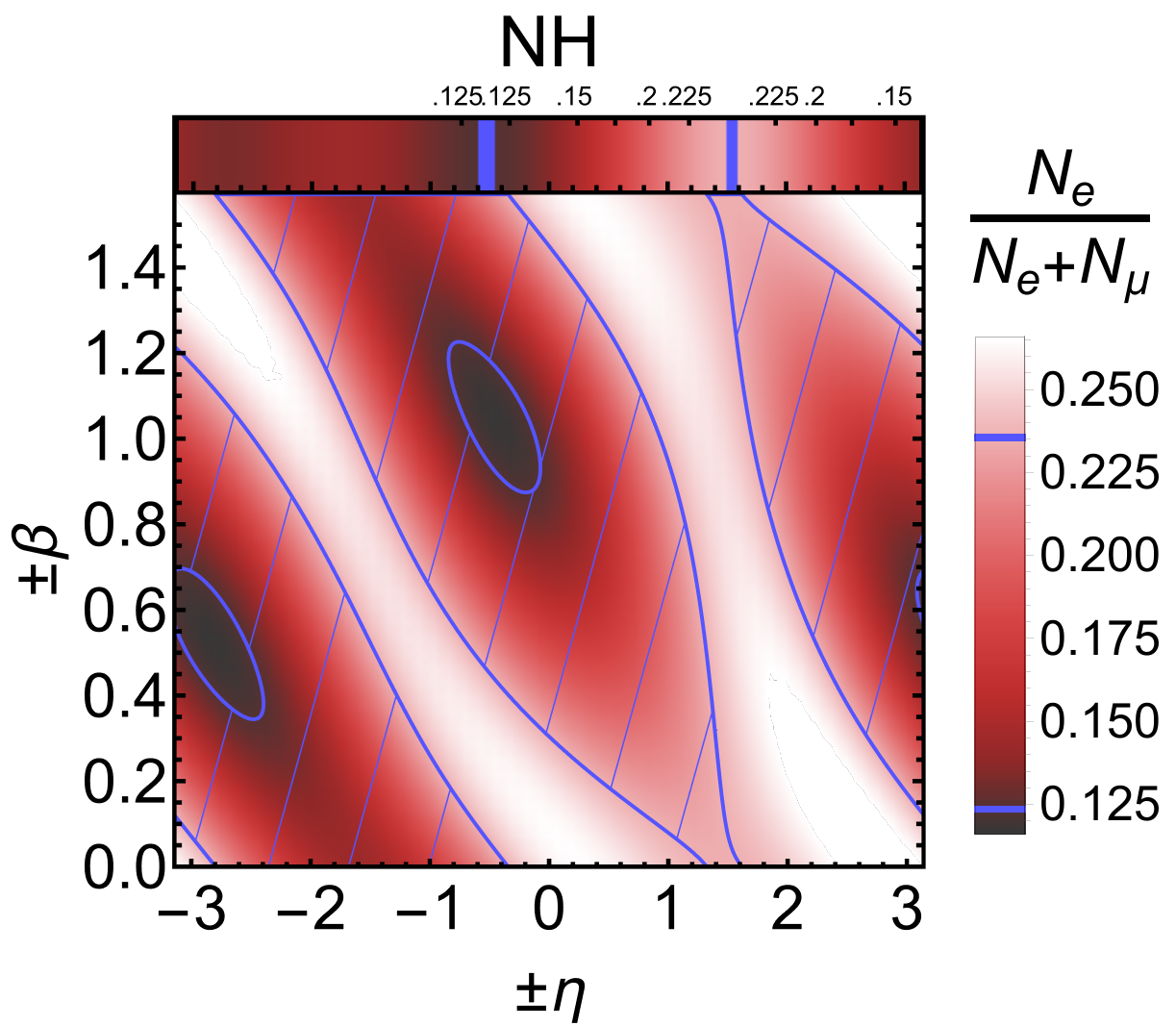}~    \includegraphics[height = 0.4\textwidth]{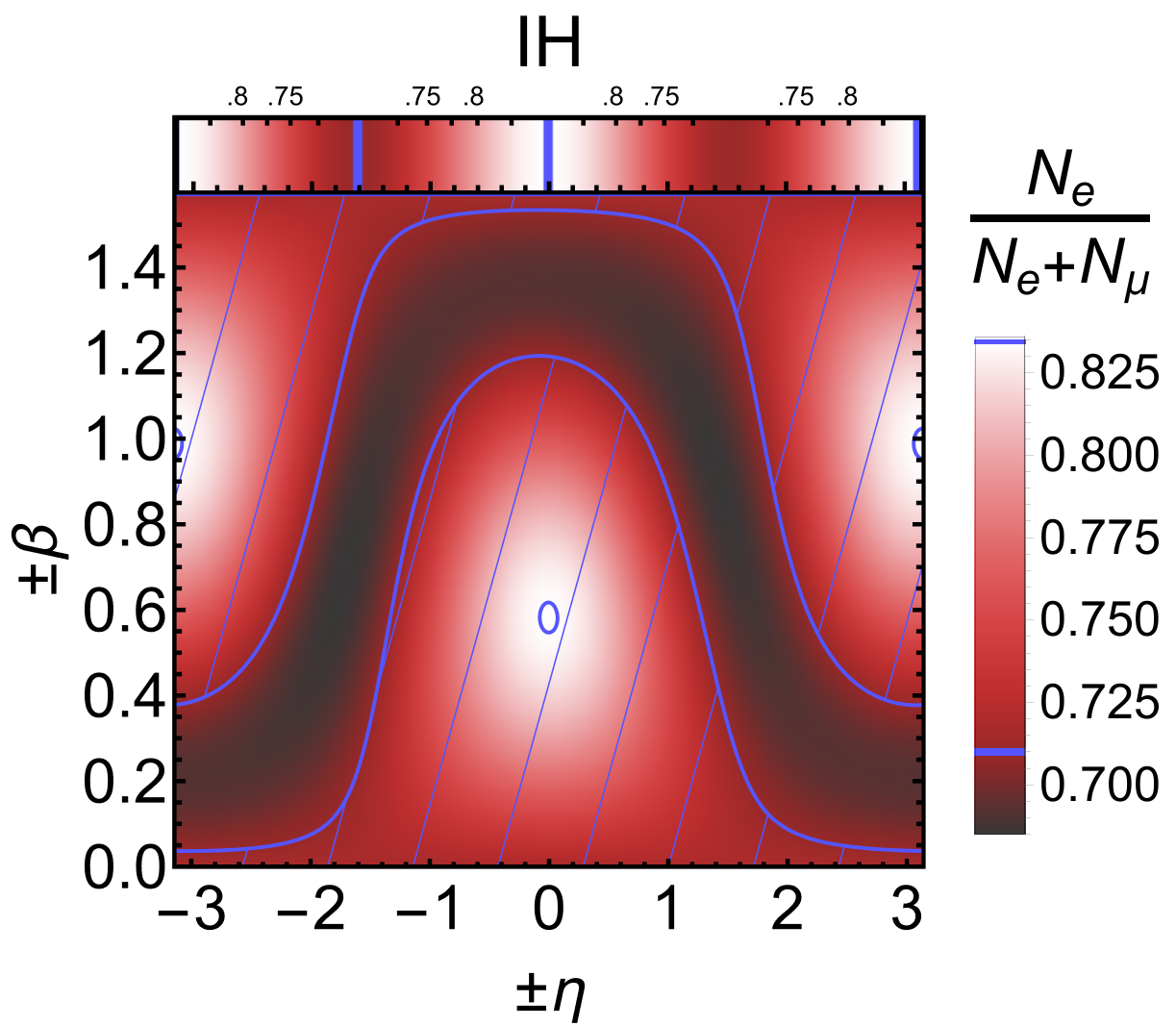}
    \caption{The fraction $N_e/(N_e+N_\mu)$ of observed decays into $e$ versus $e+\mu$  as a function of $\eta$, $\beta$ for normal (left) and inverse (right) hierarchy. HNLs with a mass around $\unit[1]{GeV}$ and only right-handed interactions are considered. Corrections due to nonzero muon mass are neglected, assuming $m_\mu \ll m_N, m_D - m_N$. The absence of $\tau$-coupled interactions allows the extraction of the flavor structure even without observing the initial lepton in the $D^\pm\to X l^\pm_\alpha N$ decay.
    The colored horizontal band on top shows the dependence on $\eta$ (bottom axis) for the coherent case, with the range of possible values (upper axis) limited by the blue bands in the colormap.
    The main plot shows the dependence of the ratio on the parameters $\beta$, $\eta$ for the decoherent case, with the dashed regions depicting the values that lie between the blue band. The coherent case can be excluded for points outside the dashed region. The plots are periodic in $\beta$, $\eta$, and the plus-minus sign corresponds to the sign in Eq.~\eqref{eq:typeIsolution}.
    }
    \label{fig:SHiP-oscillations}
\end{figure}

At \textbf{SHiP}, oscillations can lead to potentially observable position-dependent features in the kinematic distributions of decay products~\cite{Tastet:2019nqj}, if their length is of the $\unit[1]{m}$ scale, comparable to the size of the experiment. Otherwise, one has to rely directly on the event matrices in order to probe the nature of HNLs. New physics particles are produced inside the target, propagate to the decay vessel, and decay into a charged lepton and hadrons. The measurement of particle mass is possible with the kinematics of the decay products, while the absence 
of leptonic decays would indicate that the right-handed interactions are dominant. The summation over the initial lepton flavors $\alpha$ loses some information, but not as drastically as~\eqref{eq:reduced_nll} suggests. The reason is the kinematic difference between charged leptons in the range of HNL masses relevant for beam-dump experiments, namely, around a few GeV. For a concrete example, we consider HNLs with a mass of $\sim \unit[1]{GeV}$: the main production channel is decays $D^\pm_{(s)} \to X l^\pm N$ of $D$-mesons, produced in the proton scatterings. For this mass, tau-leptons are completely absent, $P_\tau = D_\tau = 0$, while electrons and muons may treated as equivalent: $P_e = P_\mu$, $D_e = D_\mu$ up to small corrections due to muon mass. Assuming that these leptons are detected with a unit efficiency, the theoretical prediction of the ratio $N_e/(N_e+N_\mu)$ between the number of events with electron/muon in the final state can be computed directly. For the decoherent case, the range of values the ratio can take is somewhat larger than that of the coherent case. Fig.~\ref{fig:SHiP-oscillations} shows the results for $N_e/(N_e + N_\mu)$; if the experimental measurement lies outside of the range defined by the blue bands (outside the dashed region), the coherent case in the minimal scenario is excluded.

\begin{figure}[t]
    \centering
    \includegraphics[width = \textwidth]{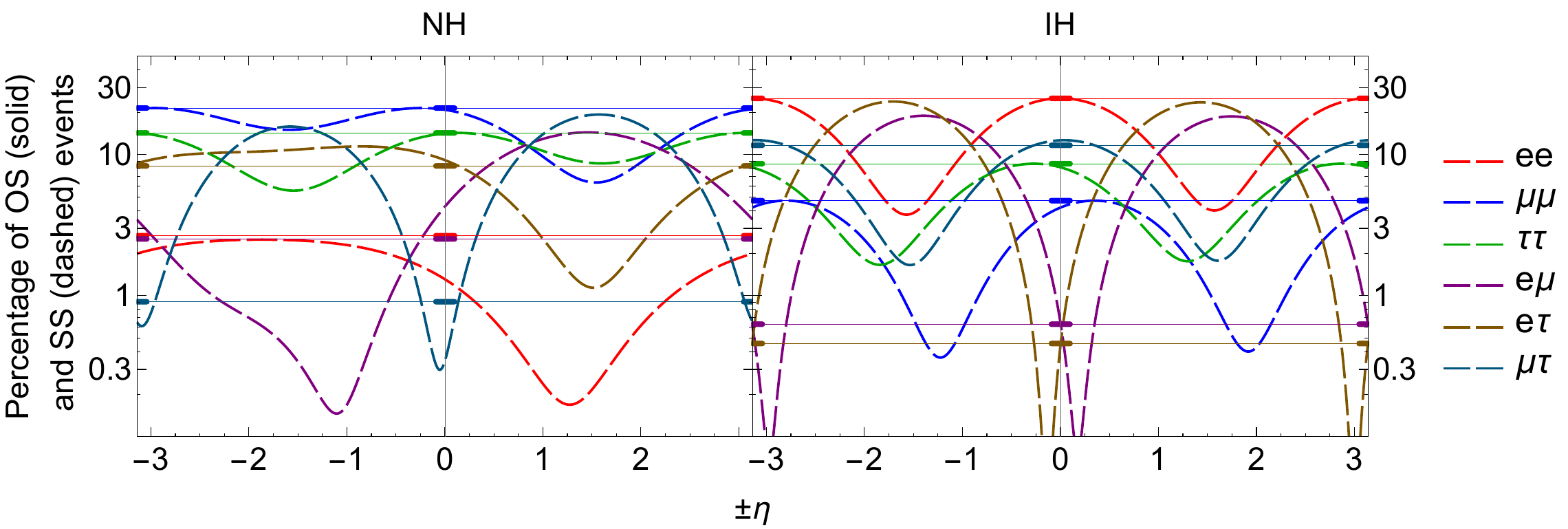}
    \caption{The relative fractions of opposite-sign (OS, solid lines) and same-sign (SS, dashed lines) events for various combinations of lepton flavors in the final state and normal (left) or inverse (right) neutrino mass hierarchy. For some values of the Majorana phase $\eta$, a conspicuous feature is the peculiar combination of a significant fraction of lepton number violating events with a much smaller amount of lepton number conserving decays. The feature is most prominent in the $\mu\tau$, $e\mu$ channel for NH, and the $e\mu$, $e\tau$ channels for IH.}
    \label{fig:SStoOS}
\end{figure}

At \textbf{hadron colliders}, the Keung-Senjanovi\'c process $pp\to X l_\alpha N \to X l_\alpha (Y l_\beta)$ is a powerful probe since both leptons can be traced, and the event matrix can be fully reconstructed. Misidentification of the initial and final lepton, which might be an issue for short-lived HNLs, does not lose information, provided that the event matrices are symmetric. With the twelve independent entries of $n_{\alpha \beta}$ and $n_{\alpha \bar \beta}$, a full fit of the model can be performed in the standard way. We focus on the relation between the lepton number violating same-sign (SS) versus conserving opposite-sign (OS) events. We assume that lepton masses can be neglected, and consider only the coherent case since, for the decoherent case, the lepton number conserving/violating events are trivially expected to occur with equal probability. The relative fractions of SS and OS to the total number of events as functions of the Majorana phase $\eta$ for different combinations of lepton flavors are shown in Fig.~\ref{fig:SStoOS}. The fractions of OS events are fixed and can be used to check the model's consistency. The SS events provide information on $\eta$.

At \textbf{lepton colliders}, the unified low-energy phenomenology described in Sec.~\ref{sec:pheno} is not valid. The analysis of LRSM at lepton colliders can be found in Ref.~\cite{Urquia-Calderon:2023dkf}. Two main differences that have to be taken into account are HNL production through the neutral $Z_R$-boson of the extended gauge group and the breakdown of the equivalence between the left- and right-handed couplings --- the adopted definition of $U^R$ loses its physical interpretation. For a concrete example, the HNL production cross-section for the left~\cite{Mikulenko:2023ezx} and right-handed interactions~\cite{Urquia-Calderon:2023dkf} have the following scalings:
\begin{equation*}
    \sigma^L \sim \frac{U_L^2}{m_W^2}, \qquad\qquad  \sigma^R(s<m_{W_R}^2) \sim \frac{s}{m_{W_R}^4} \sim \frac{U_R^2}{m_W^2} \times \left(\frac{s}{m_{W_R}^2}\right)
\end{equation*}
where the center-of-mass energy $\sqrt{s}$ is conservatively assumed to be below the $W_R$ mass. It might be possible that the LH-production dominates the RH one for the parameter space where, technically, $U_R > U_L$ for the associated couplings.  

The main process for HNLs with RH-interactions is a pair production and decay $l^+ l^- \to N_I N_J \to (l_1 X) (l_2 Y)$. The flavor and charge composition of the final leptons provides a window into the properties of HNLs. In comparison with the previous setups, the initial lepton flavor and charge are fixed and two HNLs are present, making the relation of experimental measurements to the couplings $V^R$ more convoluted. Therefore, we do not present concrete numerical results but only a qualitative analysis. The same/opposite sign signature, which refers to the charges of $l_1l_2$, still serves as a powerful probe of nontrivial dynamics in the HNL sector.  The opposite-sign events occur through double LNV or double LNC decays of the two $N$, while the same-sign events require one LNV and one LNC decay. Deviation of the SS/OS ratios from the trivial unit (Majorana/decoherent case) or zero (Dirac) values would be an indication of the coherent pair. The exact procedure for computing these ratios for the coherent case has to account for \textit{two} interferences: one coming from the production of HNLs and one coming from their decays as a single tied state. The schematic formula for the amplitude is:
\begin{align*}
    \langle l^+_\beta l^{\pm}_\gamma|&(NN\text{-mediated})|l^+_\alpha l^-_\alpha\rangle 
    \\
    &\Longrightarrow \sum_{IJ} \langle l^+_\beta l^{\pm}_\gamma|(\text{decay})|N_I N_J\rangle (C_W V_{\alpha I} V^*_{\alpha J} + C_W V^*_{\alpha I} V_{\alpha J} + C_Z \delta_{IJ})  
    \\
    & \Longrightarrow \sum_{IJ} (\tilde C_W V_{\beta I} V^{(*)}_{\gamma J} + \tilde C_W V^{(*)}_{\gamma I} V_{\beta J}) (C_W V_{\alpha I} V^*_{\alpha J} + C_W V^*_{\alpha I} V_{\alpha J} + C_Z \delta_{IJ})  
\end{align*}
where $C_W$, and $C_Z$ are kinematic factors for production through the charged and neutral current, respectively, and $\tilde C_W$ is the kinematic factor for HNL decay. These factors are independent of the lepton flavor by construction, once charge lepton masses are neglected. The notation $V^{(*)}$ refers to $V$ for the $l^+_\gamma$ and $V^{*}$ for the $l^-_\gamma$ choice. For a given model, the kinematic factors can be computed explicitly and substituted into the formula to give the probability of various combinations of final lepton states.

\subsection{Dark matter at SHiP}

\begin{figure}[t]
    \centering
    \includegraphics[width = \textwidth]{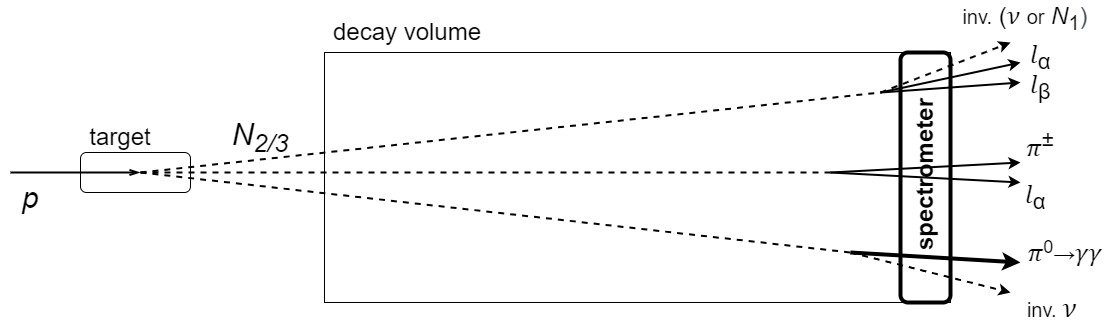}
    \caption{Schematic depiction of new physics detection at SHiP. HNLs are produced in proton-nucleus collisions in the target, move into the decay volume, and decay via one of the three main decay channels: leptonic $l\bar l+$invisible neutrino of light $N_1$, $l$CC into a charged lepton and hadrons, and NC into hadrons with total charge zero plus invisible neutrino. }
    \label{fig:SHiP_scheme}
\end{figure}

\begin{figure}[t]
    \centering
    \includegraphics[width = \textwidth]{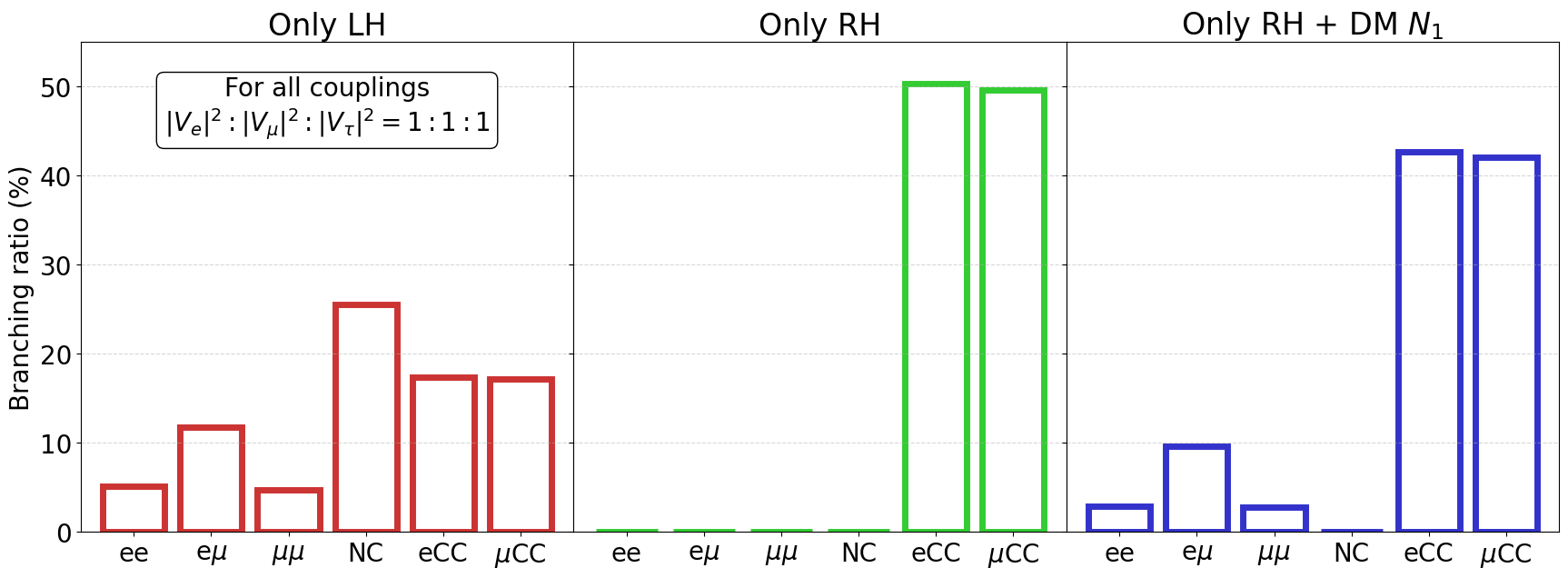}
    \caption{Branching ratios for different theoretical setups: dominant left-handed interactions (left), right-handed interactions (middle), and right-handed interaction  \textit{with} light $N_1$.}
    \label{fig:SHiP_branchings}
\end{figure}

This section aims to estimate the potential of the SHiP experiment to measure non-minimal interactions of HNLs and ultimately probe the DM candidate $N_1$. The SHiP experiment offers excellent sensitivity to long-living HNLs in both the minimal scenario~\cite{SHiP:2018xqw} and the LRSM~\cite{Castillo-Felisola:2015bha}. For the dark matter searches, the situation is different: the conventional approach~\cite{SHiP:2020noy} relies on the scattering of new physics in the Scattering and Neutrino Detector and has limited to no potential when considering the sterile neutrino. The reason is that $N_1$ interacts even weaker than the active neutrino, for which less than $10^7$ events are expected~\cite{Pastore:2020dgg}. In addition, the active neutrino events constitute a background for the new physics search, further suppressing the sensitivity. In the minimal case of left-handed interactions, the search for DM in the decays of long-living HNLs is not feasible as well: the production of the light DM is suppressed by the small mixing angle squared, one for every new species. In contrast, the right-handed interactions do not have such a problem, for the equivalent small coupling originates from the propagator of the heavy $W_R$, and thus, the probability is suppressed only once. 

The analysis is based on the branching ratios of visible decay modes, following the methodology of~\cite{Mikulenko:2023iqq, Mikulenko:2023olf}. The main idea is the following: $N_1$ is expected to be much lighter than the detectable HNL pair $N_{2/3}$ and would give rise to the decay:
\begin{equation*}
    N_{2/3} \to l_\alpha \bar l_\beta + \text{invisible}
\end{equation*} 
which mimics a similar process with a neutrino. The way this decay adjusts the observed branching ratios is, however, non-trivial. 
The overall scheme of the experiment is shown in Fig.~\ref{fig:SHiP_scheme} with the decay modes of interest: $l\bar l$, $l$CC (charged lepton and hadrons) and NC (neutrino and neutral hadrons). The $l$CC mode provides a precise measurement of the HNL mass. Simpler models without a light $N_1$, such as the commonly accepted phenomenology of HNL with left-handed interactions, would be insufficient and fail to fit the observed set of final states. 

Three basic scenarios serving as starting points are: LH-dominated case, RH-dominated case, and RH-dominated with a light $N_1$ species. The qualitative picture highlighting the differences between the three cases is shown in Fig~\ref{fig:SHiP_branchings}. The main feature of both RH cases is the absence of the NC mode. Furthermore, HNLs without the $N_1$ decay only semileptonically into $l$CC, making this case the most easily identifiable. The LH and RH+$N_1$ are more difficult to distinguish since both have the $l_\alpha l_\beta$ mode. To establish the difference, one has to rely on the NC mode, as well as on the relative ratios between $l\bar l$ and $l$CC. These three scenarios can readily be distinguished with as few as tens of events.    

Any observed experimental result must be tested compared to all possible theoretical scenarios, including the case of left and right-handed interaction strengths being of the same order. The main result of this section is a demonstration that the DM candidate can still be probed in this scenario with a sufficiently high amount of statistics.  For the demonstration, we adopt the following benchmark model: the HNLs have the mass of $\unit[1.5]{GeV}$, decay decoherently, and have couplings
\begin{align*}
    |V^L_{e}|^2 : |V^L_{\mu}|^2 : |V^L_{\tau}|^2 &= 0.11:0.22:0.67 \\
    |V^R_{e2}|^2 : |V^R_{\mu 2}|^2 : |V^R_{\tau 2}|^2 &= 0.16:0.46:0.38 \\
    |V^R_{e3}|^2 : |V^R_{\mu 3}|^2 :|V^R_{\tau 3}|^2 &= 0.16:0.46:0.38 \\
    |V^R_{e1}|^2: |V^R_{\mu 3}|^2 : |V^R_{\tau 3}|^2 &= 0.49:0.22:0.30 \\
    \mathcal R \equiv &\frac{U^2_R}{U^2_L + U^2_R}
\end{align*}
chosen to be equal for $N_2$ and $N_3$, and consistent with the parameter space of type-I seesaw of Sec.~\ref{sec:TypeIcase} (normal hierarchy). The parameter $\mathcal R$ is a free parameter that quantifies the contribution of RH interactions.

For a given $\mathcal R<1$, we want to distinguish
\begin{enumerate}
    \item the benchmark model \textit{without} light $N_1$ \textbf{versus} a single LH-interacting HNL with \textit{arbitrary} couplings $V^L$,
    \item the benchmark model \textit{with} light $N_1$ \textbf{versus} a single LH-interacting HNL with \textit{arbitrary} couplings $V^L$,
    \item the benchmark model \textit{with} the light $N_1$ \textbf{versus} a single HNL with both left and right interactions and \textit{arbitrary} couplings $V^L$, $V^R$, and $\mathcal R$, but \textit{without $N_1$}.
\end{enumerate}

\begin{figure}
    \centering
    \includegraphics[height = 0.5\textwidth]{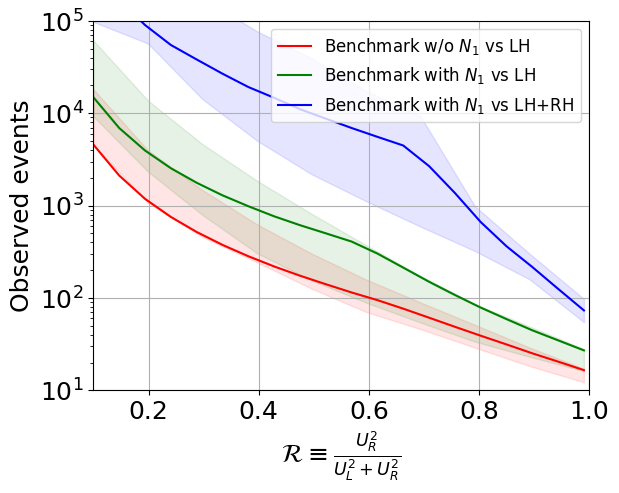}~
    \includegraphics[height = 0.5\textwidth]{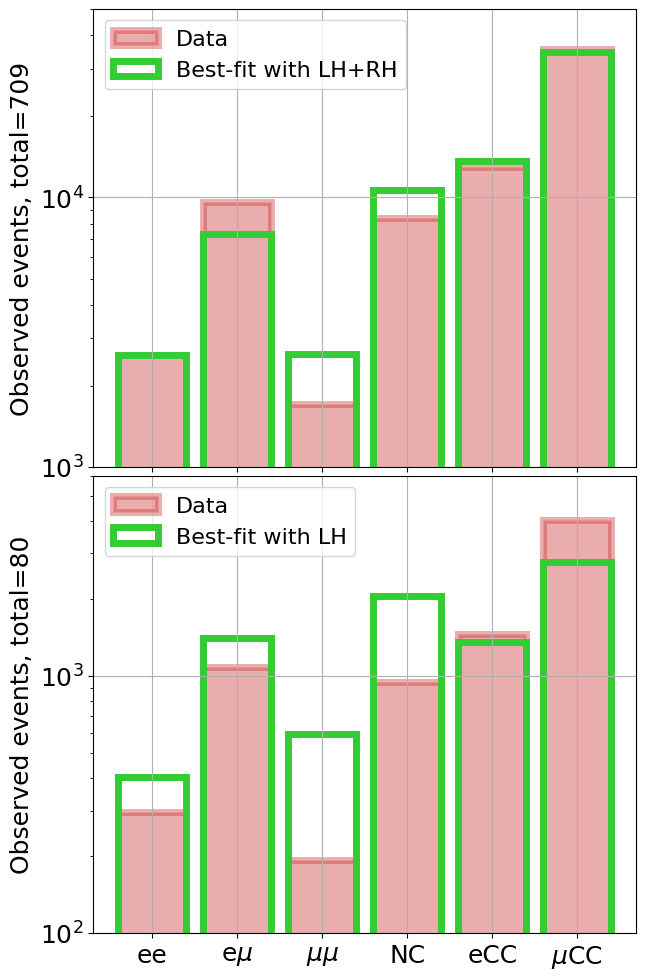}
    \caption{\textit{Left:} Number of observed events in the benchmark model, necessary to exclude the tested model versus the benchmark model at 90\% confidence level with 90\% probability. The solid lines correspond to the differentiation between the benchmark models and the tested models. The filled regions represent the variation of the results for some randomly picked combinations of couplings, consistent with Type-I seesaw case of Sec.~\ref{sec:TypeIcase}, in place of the benchmark model. \textit{Right:} an illustration of the branching ratios for the benchmark model with a DM candidate and their incompatibility with LH+RH (top) and LH (bottom) models.}
    \label{fig:NeventsDM}
\end{figure}

We compute the branching ratios for each case and use the package from~\cite{Mikulenko:2023iqq} to estimate the required number of events rejecting the tested models at 90\% confidence level with 90\% probability. We assume the most optimistic scenario of zero background and equal detection efficiency for all decay channels. The results are shown in Fig.~\ref{fig:NeventsDM}. For the dominant RH case $\mathcal R = 1$ and an existing light DM candidate, \textit{less than a hundred events} would suffice to exclude an HNL with arbitrary left and right couplings but no DM. These estimates should be treated as general lower limits and must be corrected in the proper analysis to account for the actual limitations of the detector systems: particle misidentification, different reconstruction efficiencies, and theoretical uncertainties in the branching ratios, such as those arising from the imprecise HNL mass measurement. 

It must be stressed that the third case is the most general test, assuming an HNL with arbitrary left and right couplings. In this sense, the data analysis does not have to rely in any way on the quasi-Dirac setup. Therefore, the conclusions about the possibility of probing the light candidate are not restricted to our specific model. The same analysis can be performed for an arbitrary set of couplings of the light and heavier species; the inconsistency of the signal with the model without $N_1$ will be more or less pronounced but still present, unless for some fine-tuned cases, such as complete DM-tau lepton coupling, $|V_{1 \tau}^R|= 1$.

In the case of the discovery of new physics, the observation may indicate some kind of non-minimality that requires combined left and right interactions. This would correspond to the region above the green line of Fig.~\ref{fig:NeventsDM}. In such a scenario, it would be imperative to investigate further whether the signal contains a contribution of $N_1$ decays. The blue line quantifies the minimal required sensitivity of a follow-up experiment, needed to give the decisive answer. This estimate depends on the unknown underlying theory; however, one can obtain a general reference point: In the case of a slight deviation (say, $2\sigma$) of the observation from the pure LH model, the number of required events should increase by roughly an order of magnitude.

\section{Conclusions.}

In this work, we have investigated the attractive scenario of a quasi-Dirac pair of Heavy Neutral Leptons within the Left-Right Symmetric Model. A degenerate pair can facilitate the generation of baryon asymmetry and, at the same time, be within the reach of future experimental searches without leading to too large masses of active neutrinos. Compared to the minimal case, the HNLs with the Left-Right Symmetric Model possess additional interactions through the right-handed current and have new phenomenological implications.

We analyzed the flavor structure of the right-handed HNL couplings from the full seesaw relation in the limit. In the minimal type-I case analytically, the analytical solution for the parameter space of the couplings is provided. A notable result is concrete predictions for effective right-handed couplings. These predictions serve as tangible targets for experimental validation.

Moreover, we have explored the potential impact of our findings on various experimental probes in the considered model. Our assessment of the sensitivity of FCC-ee to detect HNLs in the combined model, along with the provision of simple scaling arguments, shows that right-handed interactions can improve the sensitivity to the left-handed couplings.

We examined the properties of the quasi-Dirac HNL pair in the coherent and decoherent state. Each case offers rich phenomenology. We discussed the experimental signatures that can be employed for SHiP, FCC-ee, and future hadron and lepton colliders.

Finally, we demonstrated for the first time the potential of the recently approved SHiP experiment to probe the existence of a dark matter candidate in decays of HNLs. Our analysis suggests that, in the best-case scenario, SHiP could provide compelling evidence for dark matter with fewer than a hundred events. Our findings have become possible thanks to leveraging, on the one hand, ways to extend the new physics model by accounting for possible nonminimal interactions, with the attempt to keep the model as predictive as possible on the other hand. This highlights the necessity for further research to explore the full potential of next-generation experiments.

\section*{Acknowledgements}

I thank Kevin Alberto Urquía Calderón, Marco Drewes, Miha Nemev\v{s}ek, and Alexey Boyarsky for their comments and suggestions. This project has received funding from the European Research Council (ERC) under the European Union's Horizon 2020 research and innovation programme (GA 694896), from the NWO Physics Vrij Programme ``The Hidden Universe of Weakly Interacting Particles'', No. 680.92.18.03, which is partly financed by the Dutch Research Council NWO. Feynman diagrams have been computed with the use of \texttt{FeynCalc}~\cite{Shtabovenko:2020gxv}.

\appendix
\section{Decay $N_I \to N_J l_\alpha l_\beta$}
\label{app:decayIntoDM}

The decay width is given by
\begin{multline*} 
    \Gamma(N_I \to N_J l_\alpha \bar l_\beta) = U_R^2 \frac{G_F^2 m_{N_I}^5}{16\pi^2} \\
    \times \int^{(1-x_{J})^2}_{(x_\alpha + x_{\beta})^2} \frac{dx}{x} \sqrt{\lambda(1,x,x_J^2) \lambda(x, x_\alpha^2, x_\beta^2)} (x - x_{\alpha}^2 - x_{\beta}^2)
    \\
    \bigg((1 + x_J^2 - x) (|V^R_{\alpha I}|^2 |V^R_{\beta J}|^2 + |V^R_{\beta I}|^2 |V^R_{\alpha J}|^2)  -
    2 x_J \text{Re}[V^R_{\alpha I} V^R_{\alpha J} V^{*R}_{\beta I} V^{*R}_{\beta J}]\bigg).
\end{multline*}
Here, $x_J = m_{N_J}/m_{N_I}$, $x_\alpha = m_{l_\alpha}/m_{N_I}$, $x_\beta = m_{l_\beta}/m_{N_I}$, and
\begin{equation*}
    \lambda(a,b,c) = a^2 + b^2 + c^2 - 2ab-2ac-2bc    
\end{equation*}

\section{Square root of $2\times 2$ matrix.}

\label{app:square2x2}

Two over two matrices are sufficiently simple and an explicit formula for the square roots exists. For a complex $2\times 2$ matrix $X$, there are two pairs of square roots (except for special cases) given by~\cite{Levinger:1980}:
\begin{equation*}
    R_\pm = \frac{X + s_\pm I}{\sqrt{t + 2s_\pm}}, \qquad t = \text{Tr}\, X,\quad s_\pm = \pm \sqrt{\det X}
\end{equation*}
where $I$ is a unit matrix, and each $R_\pm$ can be further multiplied by $\pm 1$.

If the matrix $X$ is chosen to have $t \to 2|s|$, the denominator approaches zero for one of the two pairs of roots. The numerator does not vanish and the whole square root can increase indefinitely. To provide an example, a symmetric $X$ that satisfies this condition is:
\begin{equation*}
    X = \begin{pmatrix} a + b & i a \\
    i a & b - a\end{pmatrix} +O\left(\frac{1}{y}\right)
\end{equation*}
where $y\gg a, b \sim 1$ is some large scale that introduces deviations. A straightforward substitution gives a rather expected result for the large square root:\footnote{This would be the choice in the Eq.~\eqref{eq:full_symmetry} with $\tilde Y_\nu = V^* Y V^\dagger$, had we decided to follow~\cite{Nemevsek:2012iq} and adopt generalized $\mathcal C$ symmetry.}
\begin{equation*}
    R_- = O(\sqrt{y}) \begin{pmatrix}
        1 & i \\ i & -1
    \end{pmatrix} + O\left(\frac{1}{\sqrt{y}}\right)
\end{equation*}

It is crucial to note that the eigenvalues of $X$ are equal, up to the $1/\sqrt{y}$ correction, as is the case as well for the matrix of Eq.~\eqref{eq:Xmatrix}. It might be tempting to identify the eigenvalues with the neutrino masses, which are known to be different, and thus exclude the possibility that $y$ may be arbitrarily large. However, the seesaw relation is written for symmetric matrices, and the diagonalization of the matrix should be done by an orthogonal rotation $X \to O X O^T$ rather than by a unitary transformation. This operation does not leave eigenvalues invariant. Instead, the components of diagonalized $X$
\begin{equation*}
    |\lambda_\pm|^2 = (\sqrt{|a|^2+|b|^2}\pm|a|^2)^2
\end{equation*}
have enough freedom to be chosen independently in the limit of $y\to \infty$.

\bibliographystyle{JHEP}
\bibliography{bib}

\end{document}